\documentstyle[12pt,aaspp4]{article}

\slugcomment{Submitted to ApJ}

\newcommand\ep{\varepsilon}
\newcommand\th{\theta}

\newcommand\kms{km s$^{-1}$}
\newcommand\Hunits{km s$^{-1}$ Mpc$^{-1}$}
\newcommand\0{\phn}
\def\tm#1{\tablenotemark{#1}}
\newcommand\td{\tm{\dag}}
\newcommand\tdd{\tm{\ddag}}
\def\_{\discretionary{}{}{}}
\def\0{\phn}
\def\refeq#1{eq.~\ref{eq:#1}}
\def\reffig#1{Figure~\ref{fig:#1}}

\def\deg{\ifmmode ^{\circ}
	\else $^{\circ}$\fi}
\newcommand\pdeg{\ifmmode $\setbox0=\hbox{$^{\circ}$}\rlap{\hskip.11\wd0 .}$^{\circ}
	\else \setbox0=\hbox{$^{\circ}$}\rlap{\hskip.11\wd0 .}$^{\circ}$\fi}
\newdimen\sa  \newdimen\sb
\newcommand\parcs{\sa=.07em \sb=.03em
	\ifmmode $\rlap{.}$^{\scriptscriptstyle\prime\kern -\sb\prime}$\kern -\sa$
	\else \rlap{.}$^{\scriptscriptstyle\prime\kern -\sb\prime}$\kern -\sa\fi}

\begin{document}

\title{The Optical Properties of Gravitational Lens Galaxies \protect \\
as a Probe of Galaxy Structure and Evolution\footnote{Based on observations
made with the NASA/ESA Hubble Space Telescope, obtained at the
Space Telescope Science Institute, which is operated by AURA,
Inc., under NASA contract NAS5-26555.}}
\author{C.~R.~Keeton, C.~S.~Kochanek, and E.~E.~Falco}
\affil{Harvard-Smithsonian Center for Astrophysics, MS-51 \protect \\ 
       60 Garden Street \protect \\
       Cambridge, MA 02138 }
\authoremail{ckeeton@cfa.harvard.edu}
\authoremail{ckochanek@cfa.harvard.edu}
\authoremail{falco@cfa.harvard.edu}

\begin{abstract}
We combine photometry and lens modeling to study the properties
of 17 gravitational lens galaxies between $z=0.1$ and $\sim$1.
Most of the lens galaxies are passively evolving early-type
galaxies, with a few spirals.  The colors, scale lengths, and
ellipticities of lens galaxies are similar to those of the
general population of early-type galaxies, although there may
be a deficit of apparently round lens galaxies produced by the
inclination dependence of lensing cross sections.  The projected
mass distributions are aligned with the projected light
distributions to $\lesssim$10\deg, except in the presence of a
strong external tidal perturbation, suggesting that dark matter
halos have orbits that are significantly modified by interactions
with the baryonic component and are not far out of alignment
with the stars.  Lens galaxies obey image separation/lens
luminosity correlations analogous to the Faber-Jackson and
Tully-Fisher relations, which are consistent with standard
dark matter lens models.  The lens galaxy mass-to-light ratios
decrease with redshift as $d(\log M/L_B)/dz = -0.3\pm0.1$
($-0.5\pm0.1$) for $\Omega_0=1$ ($0.1$), thus providing direct
evidence of passive evolution for a sample of early-type
galaxies in low-density environments.  The evolution-corrected
mass-to-light ratios are generally larger than predicted by
constant $M/L$ dynamical models, although there is significant
scatter; with improved photometry, lens galaxy mass-to-light
ratios would better distinguish between constant $M/L$ and
dark matter models.  These conclusions are limited primarily
by the quality of lens galaxy photometry.
\end{abstract}


\section{Introduction}

Although knowledge of the distribution of galaxies in color,
luminosity, structure, and redshift is growing rapidly (e.g.\
Griffiths et al.\ 1994; Lilly et al.\ 1995; Schade et al.\ 1995;
Lin et al.\ 1996, 1997; Ellis 1997) there is still little direct
information on the masses of intermediate-redshift galaxies.  As
a result, most physical inferences about galaxy populations above
$z \sim 0.1$ (such as evolution rates) remain critically dependent
on modeling the relations between a galaxy's luminosity and its
stellar or total mass.  For nearby galaxies, masses can be
estimated by combining a dynamical model with well-defined
correlations between luminosities, scale lengths, and velocities
such as the Faber-Jackson (Faber \& Jackson 1976) and Tully-Fisher
(Tully \& Fisher 1977) relations and the fundamental plane of
elliptical galaxies (e.g.\ Djorgovski \& Davis 1987; Dressler et
al.\ 1987).  These relations have recently been extended to
$z \sim 0.5$ to study galaxy evolution (e.g.\ van Dokkum \& Franx
1996; Kelson et al.\ 1997), but the observations are difficult
because of the need for spectra with high resolution and high
signal-to-noise ratio.  Other studies, both spectroscopic (e.g.\
Bender, Ziegler \& Bruzual 1996) and photometric (e.g.\ Pahre,
Djorgovski \& de Carvalho 1996; Schade et al.\ 1996, 1997;
Stanford, Eisenhardt \& Dickinson 1997) have also found evidence
for galaxy evolution out to $z \sim 1$.  All these results apply
only to ensembles of galaxies that are in high-density, cluster
environments, where evolution may or may not be typical (see
Stanford et al.\ 1997).

There is, however, one population of galaxies whose masses can
be determined individually with high precision at any redshift,
and which are found in a wide range of environments -- gravitational
lens galaxies.  The mass enclosed by the images of a four-image
lens is determined in a model-independent way with an internal
uncertainty of only a few percent (e.g.\ Kochanek 1991a;
Wambsganss \& Paczy\'nski 1994).  The full range of cosmological
models, plausible external perturbations due to nearby objects
(e.g.\ Keeton, Kochanek \& Seljak 1997), and large-scale structure
(e.g.\ Bar-Kana 1996; Wambsganss et al.\ 1997) can change the masses
by $\lesssim$20\%.  Such precision is better than that achieved
by dynamical studies of individual nearby galaxies, and far exceeds
the precision possible for galaxies at intermediate redshifts using
dynamical methods.  Hence measuring masses via lensing allows us
to replace difficult measurements and analyses of high-precision
dynamical observations with simpler observations of source and lens
redshifts and lens galaxy photometry, making it much easier to
study physical properties such as the evolution of mass-to-light
ratios.

Lensing also offers a way to directly probe the mass distributions
of distant galaxies, and to compare them with their light
distributions.  Lens models and statistics (e.g.\ Maoz \& Rix 1993;
Kochanek 1995, 1996a; Grogin \& Narayan 1996) suggest that lens
galaxies are not well described by constant mass-to-light ratio
models and instead require dark (roughly isothermal) halos, in
agreement with results from stellar dynamics and X-ray observations
(e.g.\ Fabbiano 1989; Rix et al.\ 1997).  Lens models also suggest
that the quadrupole shapes of the mass and light distributions need
not be the same (see Keeton et al.\ 1997; Jackson, Nair \& Browne
1997), which is consistent with models of polar-ring galaxies and
X-ray galaxies indicating that dark halos can be significantly
flatter than the light (e.g.\ Sackett et al.\ 1994; Buote \&
Canizares 1994, 1996; also see the review by Sackett 1996).

We must bear in mind, however, that the lens galaxies are a biased
sample of galaxies.  First, the lens galaxy sample preferentially
selects massive galaxies, because they are more likely to lens
background objects.  For example, with an $\alpha=-1$ Schechter (1976)
luminosity function $dn/dL \propto (L/L_*)^{-1} \exp(-L/L_*)$,
and a dark matter lens model for which the lensing cross
section scales as $L$ (e.g.\ Turner, Ostriker \& Gott 1984),
the luminosity function of lens galaxies is roughly
$dn/dL \propto \exp(-L/L_*)$.  In other words, the lensing
cross section filters the divergent (by number) population of
low-luminosity galaxies out of the lens galaxy sample.  The
need to resolve the multiple images produced by a lens adds
a further bias against low-mass galaxies.  Taken together, these
mass biases mean that lens galaxies are a poor probe
of the ``faint blue galaxies'' responsible for many of the
interpretation problems in galaxy number count models (e.g.\
Ellis 1997).  The mass biases are also biases for early-type
galaxies over late-type galaxies; because of their smaller
masses, spiral galaxies are expected to account for only 10--20\%
of gravitational lenses (e.g.\ Turner et al.\ 1984; Fukugita \&
Turner 1991; Kochanek 1993a, 1996a; Maoz \& Rix 1993; Keeton \&
Kochanek 1997b).

The average lens galaxy is expected to be flatter than the average
galaxy (see Kochanek 1996b; King \& Browne 1996; Keeton et al.\
1997; Jackson et al.\ 1997), in part because the efficiency of
flattened galaxies as lenses is maximized when they are viewed
edge-on (Kochanek 1996b; Keeton \& Kochanek 1997b).  Despite
the ``inclination bias,'' the lens galaxy sample is not strongly
biased to include intrinsically flat galaxies, because
inclination-averaged lensing cross sections are almost independent
of the intrinsic axis ratio (Keeton \& Kochanek 1997b).  Finally,
optically-selected samples are biased against dusty (e.g.\
Kochanek 1991b, 1996a; Tomita 1996; Falco, Kochanek \& Mu\~noz
1997a; Malhotra, Rhoads \& Turner 1997; Perna, Loeb \& Bartelmann
1997) and bright (Kochanek 1991b, 1996a) lens galaxies; because
of these two effects the lensed quasar sample is unlikely to
contain many examples of lensing by spiral galaxies.  Statistical
models of radio-selected lenses suggest that the optically-selected
quasar lens samples may be only $50_{-20}^{+40}\%$ complete (Falco
et al.\ 1997a).  The radio-selected lenses should be a fair sample
of galaxies by mass, except for the selection against low-mass
galaxies created by the need to resolve the lensed images.  

Our goal is to assemble a preliminary survey of the optical
properties of gravitational lens galaxies, and to establish
their utility as probes of the structure and evolution of
galaxies between $z=0.1$ and 1.  Our results will be limited
by the number of objects available for study, and by the
heterogeneity and quality of the available surface photometry.
Neither of these problems is fundamental, and a determined
observational program could eliminate these restrictions.
In \S2 we gather the available data and discuss their quality.
We describe new surface photometry of Hubble Space Telescope
images, discuss photometric corrections used to convert
apparent magnitudes to rest-frame luminosities, and present
lens models to characterize lens galaxy mass distributions.
In \S3 we analyze the optical and mass structures of lens
galaxies, the colors of lens galaxies, the correlation between
image separation and lens luminosity, and lens galaxy
mass-to-light ratios and evolution.  In \S4 we present our
conclusions.

\section{Lens Galaxy Data}

There are now $\sim$30 known multiply-imaged gravitational
lenses.\footnote{For a summary, see Keeton \& Kochanek (1996) or
http://{\_}cfa-www.harvard.edu/{\_}glensdata.}
In some lenses the sources are so bright that no lens galaxy has
been seen (e.g.\ the doubles 1208$+$1011 and J03.13, and the quad
H~1413$+$117), and in others we have no way to determine whether
the optical emission is due to the source or to the lens (e.g.\
B~1938$+$666 and PKS~1830$-$211).  There are, however, 21 lenses
where the lens galaxy has been observed, although the quality
of the data varies.  We have excluded the lens Q~2237$+$0305
(Huchra et al.\ 1985), which consists of four images of a quasar
produced by the bulge of a nearby ($z = 0.04$) spiral galaxy,
because it is difficult to find a luminosity estimate for the
bulge that plays the same role as aperture or total magnitudes
for the other lens galaxies.  We also excluded the lens
MG~2016$+$112 (Lawrence et al.\ 1984) because it seems to require
two lens galaxies at different redshifts and is not well
understood (see Nair \& Garrett 1997).  Finally, we neglected
the lenses FSC~10214$+$4724 (Eisenhardt et al.\ 1996) and
HE~1104$-$1805 (Wisotski et al.\ 1993; Courbin, Lidman \&
Magain 1997a).  Thus we are left with 17 lens galaxies, whose
optical properties we summarize in Tables~1--4.  We are
interested in the lens geometries and redshifts (Table~1),
the lens galaxy structural parameters (effective radius,
ellipticity, and position angle; Table~2), as well as the lens
galaxy color (Table~2), total integrated magnitude (Table~3),
and magnitude inside the critical radius (Table~4).

Only a few lens galaxies have structural parameters reported in
the literature.  More have published photometry, but because the
defining aperture is usually unspecified it is difficult to
interpret the data.  Many of the lenses, however, have been
observed with the Hubble Space Telescope (HST); with these
observations we were able to make a systematic analysis of lens
galaxy structural parameters and magnitudes.  We used our own HST
data together with data from the HST public archive (see Table~2
for references).  For the analysis, we combined each set of
cosmic-ray split images by standard procedures and constructed
a synthetic PSF with Tiny Tim (v.\ 4.3, Krist 1997).  We then
simultaneously modeled the images as a combination of point
sources or Gaussians (for the lensed images) and elliptical
de Vaucouleurs profiles (for the galaxies).  Each model was
computed on a three-times oversampled grid relative to the PSF,
which itself was modeled on a two-times oversampled grid relative
to the images.  The model was optimized to derive the structural
parameters of the lens galaxy, which could then be used to compute
total and aperture magnitudes.  Note that we quote HST instrumental
magnitudes rather than converting them to ``standard'' filters
(such as $BV\!RI$).  Finally, we used the full covariance matrix
of the model fit to derive formal statistical uncertainties
(after rescaling the nominal $\chi^2$ to be unity per degree of
freedom).  Our procedures are similar to those used by the Medium
Deep Survey (e.g.\ Ratnatunga et al.\ 1995), and for the two
lenses found in the MDS survey (HST~12531$-$2914 and
HST~14176$+$5226) we derive statistically compatible results.
We use only de Vaucouleurs fits to the galaxies because isolated
exponential disks failed to provide comparably good fits, and
the complication of disk+bulge models is unwarranted at present.
Several of the galaxies require special discussion:
\begin{itemize}

\item B~0218$+$357: We could not robustly separate the quasar
images from the galaxy in the existing data.  We used the results
from similar modeling by B.~McLeod (1997, private communication).

\item Q~0142$-$100 and BRI~0952$-$0115: In each case the galaxy 
properties are sensitive to the treatment of the bright quasar
images that bracket the galaxy, and the formal statistical
uncertainties underestimate the true systematic errors in the
profile fits.  A blind deconvolution is needed to construct an
accurate model of the true PSF simultaneously with the model of
the image.

\item Q~0957$+$561: The galaxy parameters are robust, but as in
Q~0142$-$100 and BRI~0952$-$0115 the model would benefit from a
blind deconvolution, particularly to determine the properties of
the low surface brightness (lensed) features discovered by
Bernstein et al.\ (1997).

\item MG~1131$+$0456: There appears to be a very low surface
brightness optical ring surrounding the lens galaxy.  An optical
ring was suggested by Hammer et al.\ (1991), but their detection
is far too bright to match the HST images and must have been
created by an error in modeling the galaxy profile and the
ground-based PSF.

\item B~1600$+$434: Both the HST $V$ and $I$ images have little
flux from the galaxy, although the mean structure is readily
apparent if we subtract the lensed images and convolve the residual
image with a Gaussian to increase the visibility of the low surface
brightness galaxy.  There are two peaks to the extended emission,
one between the two lensed images, and one underneath the fainter
radio image (see Jackson et al.\ 1997).  The low signal-to-noise
ratio and the complicated structure prevent any sensible model
parameter determinations.  The shape, orientation, and color were
estimated from aperture photometry.

\item B~1608$+$656: In addition to the obvious extended lensed
emission (Jackson et al.\ 1997), there appears to be a faint ring
surrounding the entire galaxy, particularly in the $I$ image.
Despite its faintness, the emission strongly affects the profile
fits, and a reliable determination of the properties of the lens
galaxy will require using a lens model to properly subtract the
optical ring.  In addition, the lens galaxy shows a complicated
structure that may indicate that the lensing mass comprises two
galaxies or a single galaxy cut by a dust lane (Jackson et al.\
1997).

\item B~1933$+$503: As with B~1600$+$434, the rough appearance of
the galaxy is readily apparent after convolving the images with a
Gaussian to enhance the low surface brightness features.  However,
there is far too little flux from the galaxy to do more than crudely
estimate an ellipticity, position angle, and aperture magnitude.
  
\end{itemize}
Where we could not robustly extract all the desired data, we either
made estimates and enclosed them in parentheses in Tables~2--4 to
flag them as systematically suspect, or omitted them altogether if
we judged them too unreliable.  In most cases the effective radius
estimate and hence the total magnitude are the most uncertain
quantities, while the ellipticity, position angle, color, and
aperture magnitude are relatively reliable.

There are several corrections that should be applied to the
magnitudes.  First, Table~1 gives $E(B-V)$ values for Galactic
extinction, estimated from the survey of Stark et al.\ (1992).
The large foreground extinctions for several lens galaxies
(notably the radio-selected lenses MG~0414$+$0534, B~0712$+$472,
and B~1933$+$503) produce significant color and magnitude
corrections that have not previously been taken into account.
We corrected colors and magnitudes for Galactic extinction using
extinction coefficients computed with $R_V=3.3$ and the Cardelli,
Clayton \& Mathis (1989) model for the extinction curve.  Note,
however, that Tables~2--4 give instrumental colors and magnitudes
without extinction corrections.

Second, as indicated by Tables~3--4, the existing observations use
a very heterogeneous set of filters.  Comparing the observations to
one another and to studies of local galaxies requires converting to
a standard wavelength, which can be done by applying photometric
corrections including color, $K$, and evolutionary corrections.
We computed the corrections using the spectral evolution models
of Bruzual \& Charlot (1993), with the Kurucz (1979) model of
$\alpha$Lyrae to define the standard magnitude zero points.
Fukugita, Shimasaku \& Ichikawa (1995) discuss the normalizations
of a wide variety of photometric systems, and Holtzman et al.\
(1995) discuss the normalization of the HST/WFPC2 photometric
bands.  We checked our results against those of Guiderdoni \&
Rocca-Volmerange (1987, 1988) and Poggianti (1997), both of
which describe the standard technique.

The photometric corrections introduce several systematic uncertainties
into the converted magnitudes:
(1) the choice of the galaxy evolution model, in particular the
galaxy formation redshift $z_f$ and the initial mass function
(IMF);
(2) the cosmological model, which enters the models through the
galaxy age as a function of redshift;
(3) the star formation rates used to describe different
morphological types, which we take from Guiderdoni \& Rocca-Volmerange
(1988); and
(4) the synthetic filters used to simulate real photometric
filters.
We take as our canonical models a Salpeter (1955) IMF for early-type
galaxies and a Scalo (1986) IMF for spiral galaxies, with a formation
redshift $z_f=15$ for all galaxies.  We use $H_0=50$ \Hunits\ and
consider several discrete values for $\Omega_0$ and $\lambda_0$.  We
have not attempted to assign formal uncertainties to the photometric
corrections, but based on \S3.2 below we estimate that the systematic
uncertainties (other than from the cosmological model) are a few tenths
of a magnitude.

Finally, we characterize the mass distributions of the lens galaxies
using dark matter dominated, singular isothermal ellipsoid (SIE) lens
models (see Kassiola \& Kovner 1993; Kormann, Schneider \& Bartelmann
1994ab; Kochanek 1996b; Keeton et al.\ 1997).  We focus on these models
because galaxies generally appear to be singular (e.g.\ Gebhardt et
al.\ 1996) and because stellar dynamical models (e.g.\ Rix et al.\
1997), X-ray galaxies (e.g.\ Fabbiano 1989), and lens models and
statistics (e.g.\ Maoz \& Rix 1993; Kochanek 1995, 1996a; Grogin \&
Narayan 1996) generally prefer an isothermal profile.  Using an
ellipsoidal mass distribution allows us to probe the shape of the
mass distribution.  We use a surface mass density of the form
\begin{equation}
	{2\Sigma \over \Sigma_{cr}} =
		{b \over r \sqrt{1+\ep\cos2(\th-\th_0)}}
\end{equation}
where $b$ is the critical radius, $\ep$ an ellipticity parameter
related to the axis ratio $q$ by $q^2 = (1-\ep)/(1+\ep)$, and
$\th_0$ is the orientation angle of the projected ellipsoid quoted
as a standard major-axis position angle (measured North through
East).  The critical surface density for lensing is
$\Sigma_{cr} = 2.34 h^{-1} (D_l D_s / 2 r_H D_{ls}) \times 10^{11}\,
M_\odot/ \mbox{arcsec}^2$, where $r_H = c/H_0$ is the Hubble radius
and $D_l$, $D_s$, and $D_{ls}$ are angular diameter distances to
the lens, to the source, and from the lens to the source,
respectively.  Formulas for the deflection and magnification of
the SIE model are given in Kassiola \& Kovner (1993), Kormann et
al.\ (1994a), and Keeton \& Kochanek (1997b).  In addition to the
SIE model, another usefule isothermal model is a singular isothermal
sphere plus external shear (SIS+shear) model where the ellipsoidal
mass distribution is replaced by the combination of a spherical
mass distribution and an external tidal perturbation (e.g.\ Falco,
Gorenstein \& Shapiro 1985; Gorenstein, Shapiro \& Falco 1988;
Kochanek 1991a).  For completeness we consider both the
SIE and SIS+shear models, and we summarize the model results in
Table~5.  Note that for the table we have converted the
parameter $\ep$ to the ellipticity $e$ using $e = 1 - q$ and
$q = (1-\ep)^{1/2}/(1+\ep)^{1/2}$.

The models we consider have a single source of shear -- an
ellipsoidal galaxy or an external shear -- and Table~5 shows that
they generally cannot give a good fit.  For many lenses, obtaining
a good $\chi^2$ requires a model with two independent shears,
such as an ellipsoidal galaxy with an external shear (e.g.\ Keeton
et al.\ 1997; Witt \& Mao 1997), and in at least four cases the
external shear can be attributed to a significant tidal perturbation
from a nearby cluster, group, or galaxy
(MG~0751$+$2716, Leh\'ar et al.\ 1997a;
Q~0957$+$561, e.g.\ Young et al.\ 1980, Grogin \& Narayan 1996;
PG~1115$+$080, e.g.\ Keeton \& Kochanek 1997a, Kundi\'c et al.\
1997a, Schechter et al.\ 1997, Tonry 1997;
B~1422$+$231, e.g.\ Hogg \& Blandford 1994, Kormann et al.\
1994b, Kundi\'c et al.\ 1997b, Tonry 1997).
Nevertheless, the single-shear models are useful because they
provide fundamental information about the average monopole and
quadrupole of the mass.  They give a position angle for the mean
shear axis that is largely independent both of the mass profile
and of the nature of the shear.  They also give a robust mass
estimate, namely the mass within the critical radius $M =
\pi b^2 \Sigma_{cr}$, that is essentially model-independent
(e.g.\ Kochanek 1991a; Wambsganss \& Paczy\'nski 1994) and
has internal uncertainties of only a few percent (due to
uncertainties in $b$).  There is, in addition, a $\lesssim$10\%
mass uncertainty due to potential fluctuations from large-scale
structure (e.g.\ Bar-Kana 1996; Wambsganss et al.\ 1997) and to
perturbations from nearby galaxies or groups (e.g.\ Keeton et
al.\ 1997), as well as a small dependence ($\lesssim$10\%) on
the cosmological model through $\Sigma_{cr}$.

\section{Results}

We now combine the optical data and the lens models to survey the
properties of lens galaxies.  We consider the structures of the
mass and light distributions, the colors of lens galaxies, the
correlation between lens galaxy luminosity and image separation,
and the mass-to-light ratios and evolution of the lens galaxies.

\subsection{Lens galaxy structure}

We can ask two questions about the structure of lens galaxies.
First, because lensing makes it possible to directly probe the
mass distribution of lens galaxies, we can ask how the mass
and light distributions compare in individual galaxies.  We
know that their profiles differ from each other, because in
early-type galaxies the light roughly follows a de Vaucouleurs
(1948) $r^{1/4}$ law, while the mass must be distributed in a
more extended halo that is roughly isothermal (e.g.\ Fabbiano
1989; Maoz \& Rix 1993; Kochanek 1995, 1996a; Grogin \& Narayan
1996; Rix et al.\ 1997), but we can ask how the quadrupole
structures compare by studying the shapes and orientations of
the models and of the light distributions (also see Keeton et
al.\ 1997; Jackson et al.\ 1997).  Second, we can compare the
distribution of optical properties for the lens galaxy sample
and the general population of early-type galaxies, and thus
examine lensing selection effects and biases.

\begin{figure}
	\plottwo{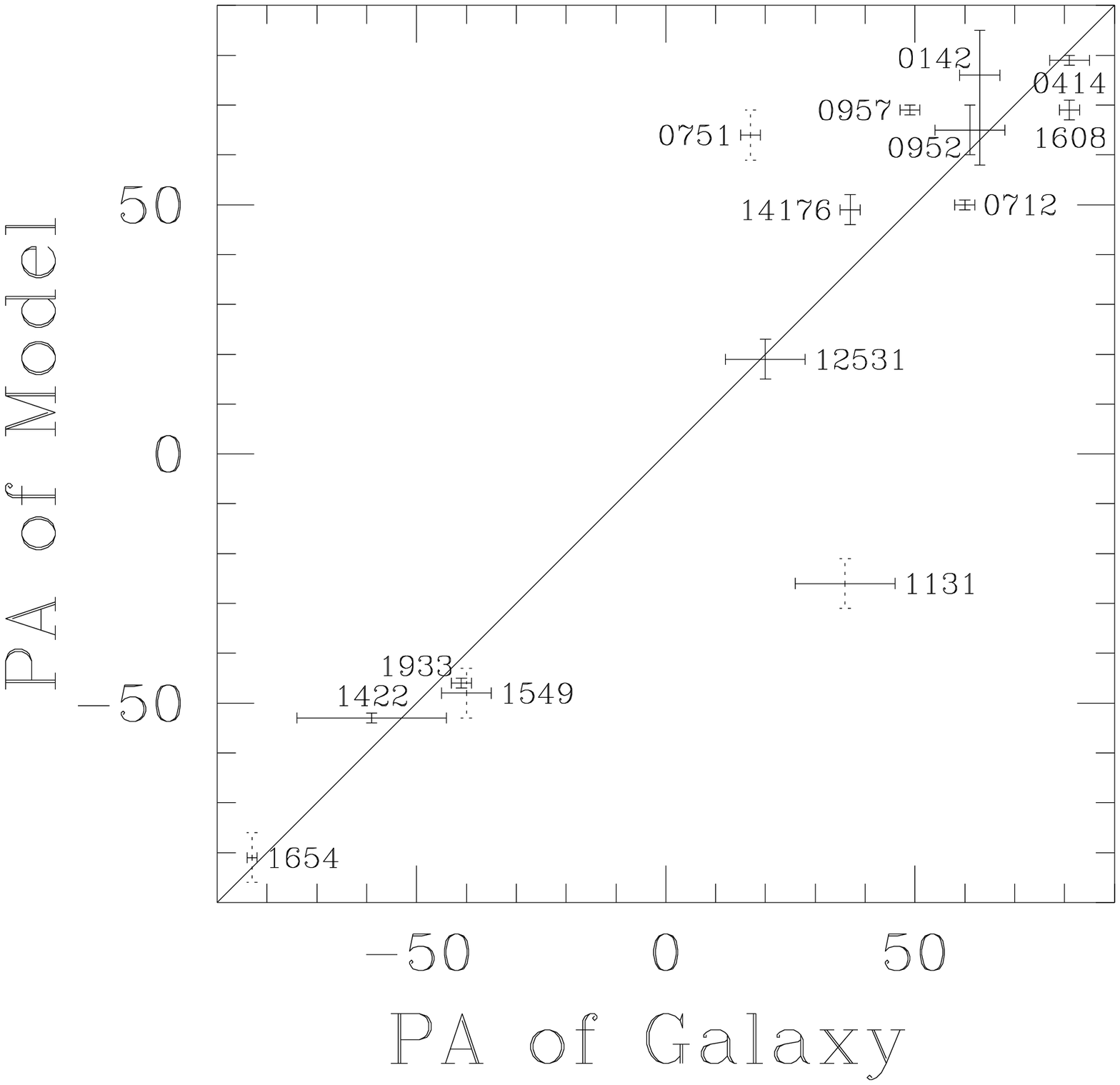}{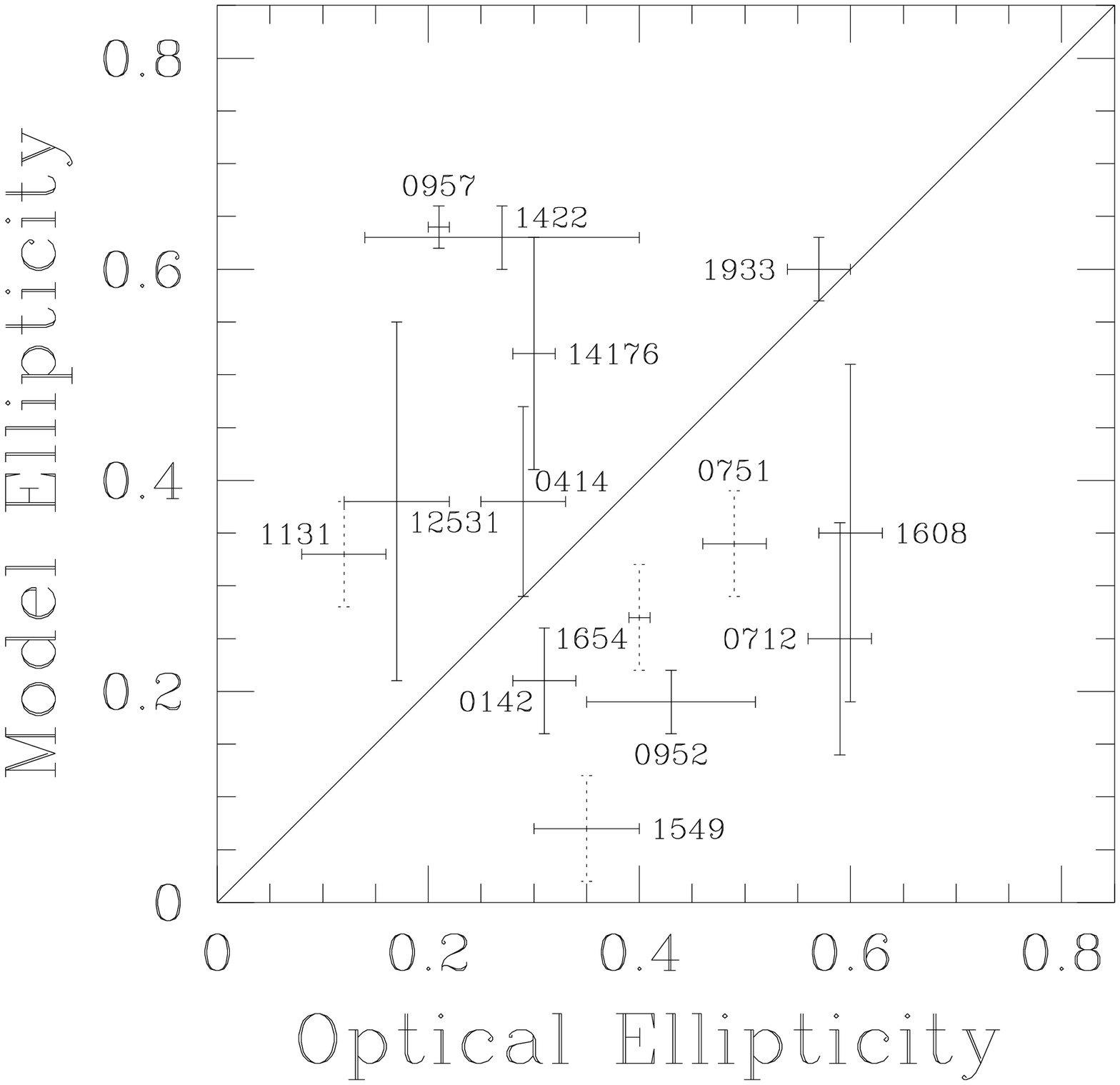}
	\caption{
Observed and model PAs and ellipticities, from Tables~2 and 5.
For model results with no formal error bars, we use error bars
of $5\deg$ in the PA and $0.05$ in $e$ and indicate them by
dotted lines.
}\label{fig:shapes}
\end{figure}

\reffig{shapes} compares the optical and model major-axis PAs of
the lens galaxies.  With a few exceptions the PAs are the same
to $\lesssim$10\deg, i.e.\ either consistent or different by
$\lesssim$2--3$\sigma$.  Two exceptions are MG~0751$+$2716
(Leh\'ar et al.\ 1997) and Q~0957$+$561 (Young et al.\ 1980),
but in these systems the lens galaxy is part of a cluster or
group that contributes to the lensing, so it is not surprising
that the average quadrupole of the mass is not aligned with the
galaxy's light.  In the remaining outlier, MG~1131$+$0456, the
PA of the model closely matches that expected if two nearby
galaxies produce a tidal perturbation (Chen, Kochanek \& Hewitt
1995).  The models appear, then, to be consistent with the
hypothesis that in the absence of external tidal perturbations
the projected mass is aligned with the projected light, and
conversely that large misalignments signal the presence of
external tidal perturbations.  Note, however, that the
alignment of the galaxy with the shear in B~1422$+$231 (Hogg
\& Blandford 1994; Kormann et al.\ 1994b; Impey et al.\ 1996;
Keeton et al.\ 1997) reminds us that even strong tidal
perturbations need not produce large misalignments.  If all
large misalignments are attributable to external tides, then
halos cannot be far out of equilibrium.  In addition, halos
probably cannot have the flat, nearly prolate shapes predicted
by N-body simulations of dissipationless collapse (e.g.\
Dubinski \& Carlberg 1991; Warren et al.\ 1992) because such
halos combined with modestly triaxial luminous galaxies
inferred from kinematic misalignment studies (e.g.\ Franx,
Illingworth \& de Zeeuw 1991) produce misalignments between
the major axes of the luminous galaxy and the dark halo of
$\langle\psi\rangle \simeq 16\deg \pm 19\deg$ (Romanowsky \&
Kochanek 1997).

\reffig{shapes} also compares the optical and model ellipticities.
Several lenses with large tidal perturbations (Q~0957$+$561 and
B~1422$+$231) clearly stand out by requiring models significantly
flatter than the light, although one (MG~0751$+$2716) does not.
The remaining lenses show no strong correlation between the optical
and model ellipticities.  The lack of a correlation is certainly
consistent with the local results that mass and luminous axis
ratios can differ (e.g.\ Sackett et al.\ 1994; Buote \& Canizares
1994, 1996; see the review by Sackett 1996).  A quantitative
interpretation of \reffig{shapes} is not straightforward, however,
because the model ellipticity depends on the mass profile, with
steeper density profiles requiring larger ellipticities.  Still,
it appears that galaxy mass distributions are not intrinsically
very flat, which is consistent with the fact that we do not see
any lenses with the ``disk'' image geometry -- 2 or 3 images off
to one side of the lens galaxy center and bracketing the projected
disk -- that are associated with highly flattened mass distributions
(see Keeton \& Kochanek 1997b).

\begin{figure}
	\plotone{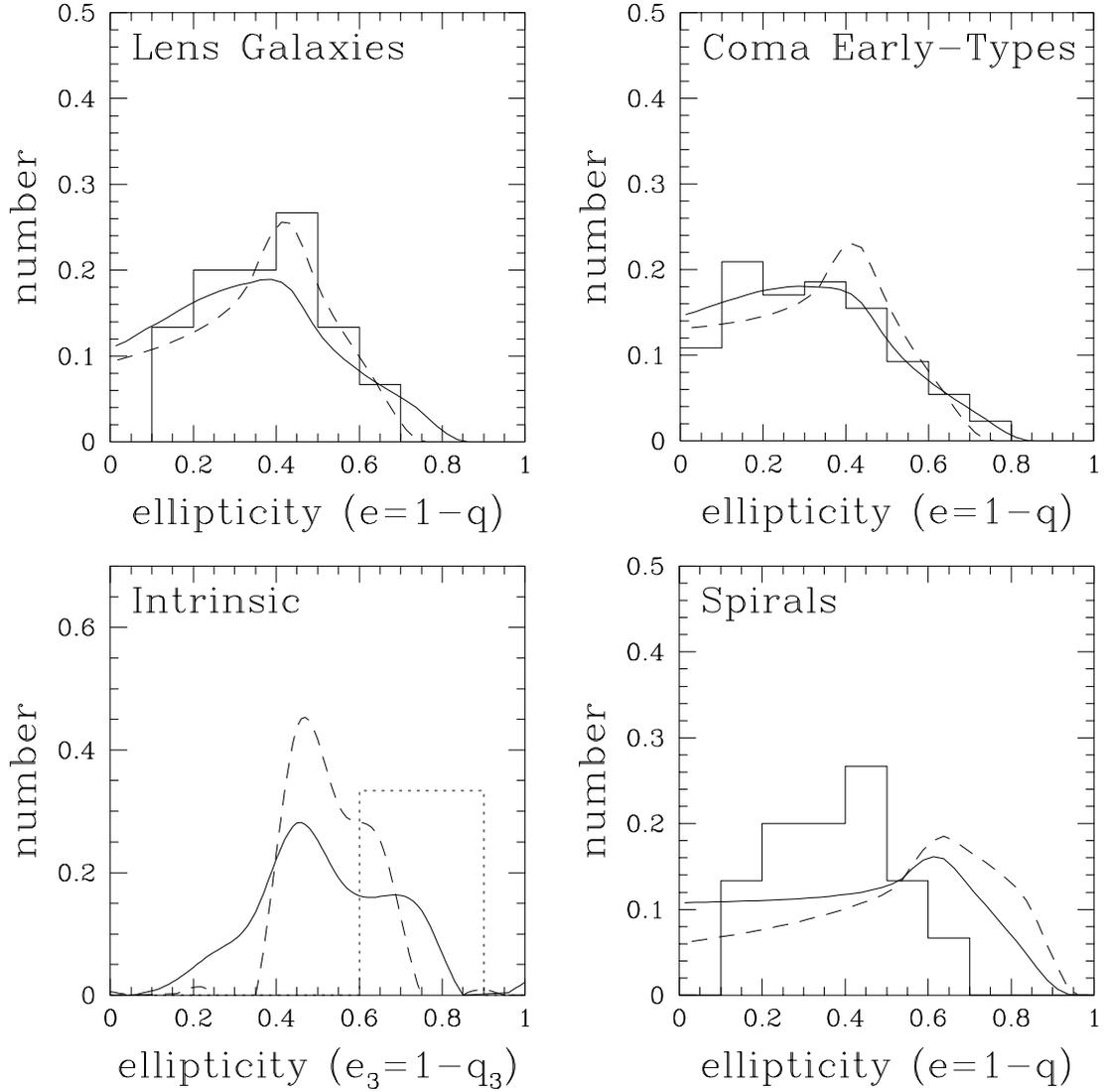}
	\caption{
Optical ellipticity distribution of lens galaxies and of a sample of
early-type galaxies in Coma (J{\o}rgensen \& Franx 1994), where $q$
is the projected axis ratio and $q_3$ is the intrinsic axis ratio.
The histograms in the upper left and upper right show the lens
galaxy and Coma samples, respectively.  In the lower left, the
solid and dashed curves show deprojections of the Coma and lens
galaxy samples, respectively, assuming galaxies are oblate.  In
the upper left and right, the curves show reprojections of the
deprojected distributions.  Note that the lensing projection and
deprojection take into account the effects of both inclination
bias (Keeton \& Kochanek 1997b) and magnification bias (Falco et
al.\ 1997a).  Finally, the lower right shows the effects expected
for spirals.  Specifically, an (assumed) intrinsic distribution
given by the dotted line in the lower left projects to the curves
in the lower right, where the solid curve is a simple projection
and the dashed curve is a lensing projection.  The lens galaxy
histogram is included for comparison.
}\label{fig:elldist}
\end{figure}

\reffig{elldist} compares the optical axis ratios of the lens
galaxies to a sample of early-type galaxies in Coma from J{\o}rgensen
\& Franx (1994).  Visually, there appears to be a deficit of both
round and flat lens galaxies compared to the Coma sample, both for
the optical quasar and radio lens samples.  A Kolmogorov-Smirnov
(K--S) test of whether the distributions are identical gives a
probability of 32\%, which indicates that any differences between
the two samples are statistically marginal.  The observed numbers
of 4-image lenses also suggest that on average lens galaxies are
flatter than observed galaxies (Kochanek 1996b; King \& Browne 1996;
Keeton et al.\ 1997), but this result is significant only at the
$1\sigma$ confidence level (Kochanek 1996b).  Thus at present the
statistical evidence that lens galaxies are significantly flatter
than regular galaxies is weak.  Such an effect could be explained
naturally by the lensing inclination bias, which makes flattened
lens galaxies likely to be seen edge-on rather than face-on (Keeton
et al.\ 1997; Keeton \& Kochanek 1997b) and predicts a $\sim$30\%
deficit of $e=0$ lens galaxies.  This effect is not strong for very
flat galaxies, because optically flat galaxies such as spirals
require a significantly rounder dark halo for stability (Ostriker
\& Peebles 1973).  Given that there is no significant difference
in the ellipticity distributions of the optical- and radio-selected
samples, and that there are few highly flattened systems, it is
unlikely that there is a significant contribution from spiral
galaxies independent of the inclination bias.

\begin{figure}
	\plotone{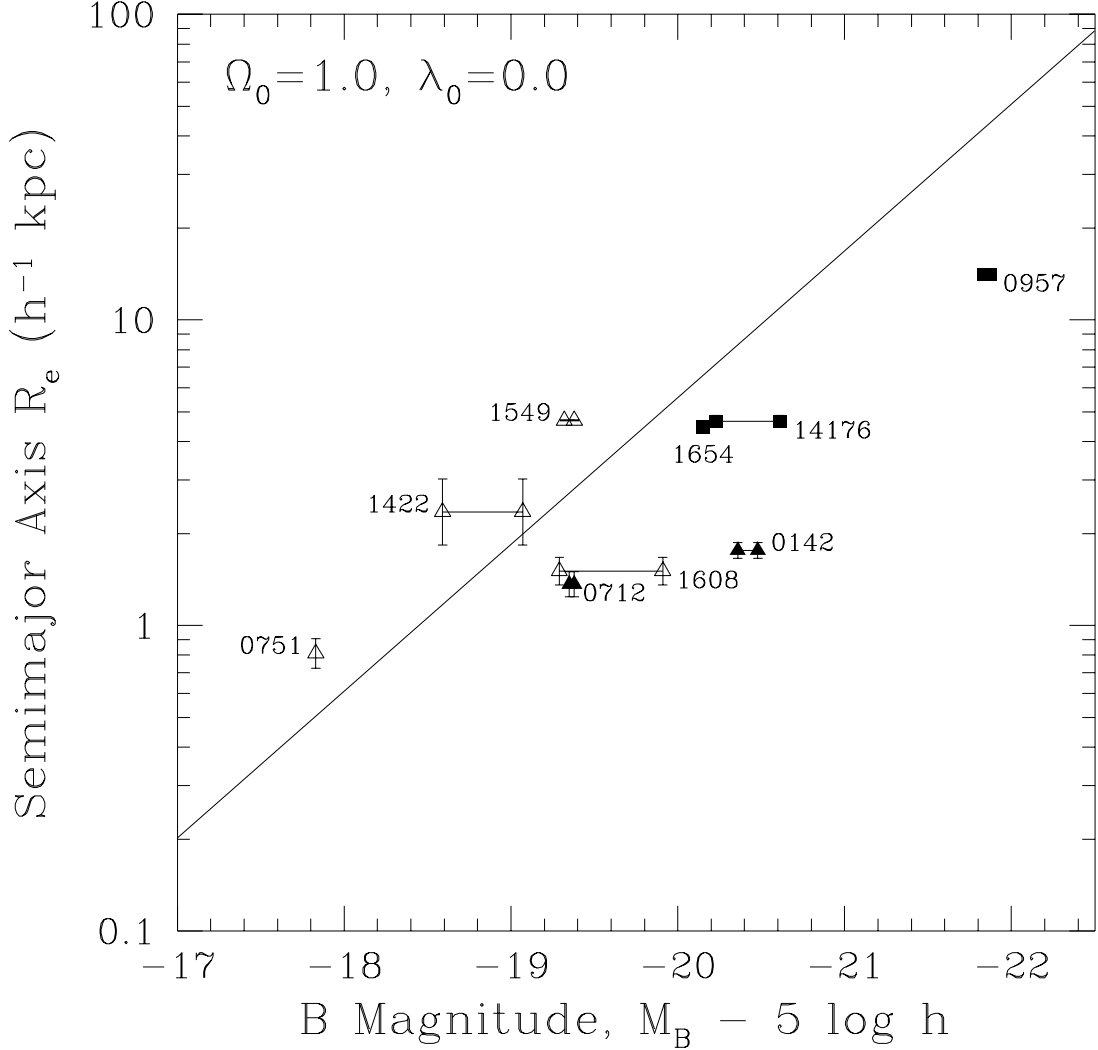}
	\caption{
Luminosities and scale lengths for lens galaxies in an $\Omega_0=1$
cosmology.  Filled (open) symbols denote luminosities computed from
total (aperture) magnitudes combined with color, $K$, evolution,
and Galactic extinction corrections.  Points connected by a horizontal
line denote magnitudes from multiple passbands.  Squares (triangles)
denote robust (uncertain) estimates for $R_e$.  The solid line shows the
correlation for nearby early-type galaxies, $R_e/R_{e*} = (L/L_*)^a$
with $R_{e*}= (4\pm1)h^{-1}$ kpc and $a=1.2\pm0.2$ (e.g.\ Kormendy \&
Djorgovski 1989; Rix 1991).
}\label{fig:re}
\end{figure}

Finally, \reffig{re} shows the relation between lens galaxy effective
radius and luminosity (for an $\Omega_0=1$ cosmology), compared to
the correlation for nearby early-types $R_e / R_{e*} = (L/L_*)^a$
with $R_{e*} = (4\pm1)h^{-1}$ kpc and $a=1.2\pm0.2$ (e.g.\ Kormendy
\& Djorgovski 1989; Rix 1991).  The lens galaxies seem to follow the
trend, although the dispersion is large.  Also, the result is not
very robust because the effective radii (and hence also the total
magnitudes) are the most uncertain of the optical parameters, and
because several of the lens galaxies (MG~0414$+$0534, MG~1131$+$0456,
and HST 12531$-$2914) that do have good estimates of $R_e$ lack lens
redshifts.

\subsection{Lens galaxy colors}

\begin{figure}
	\plotone{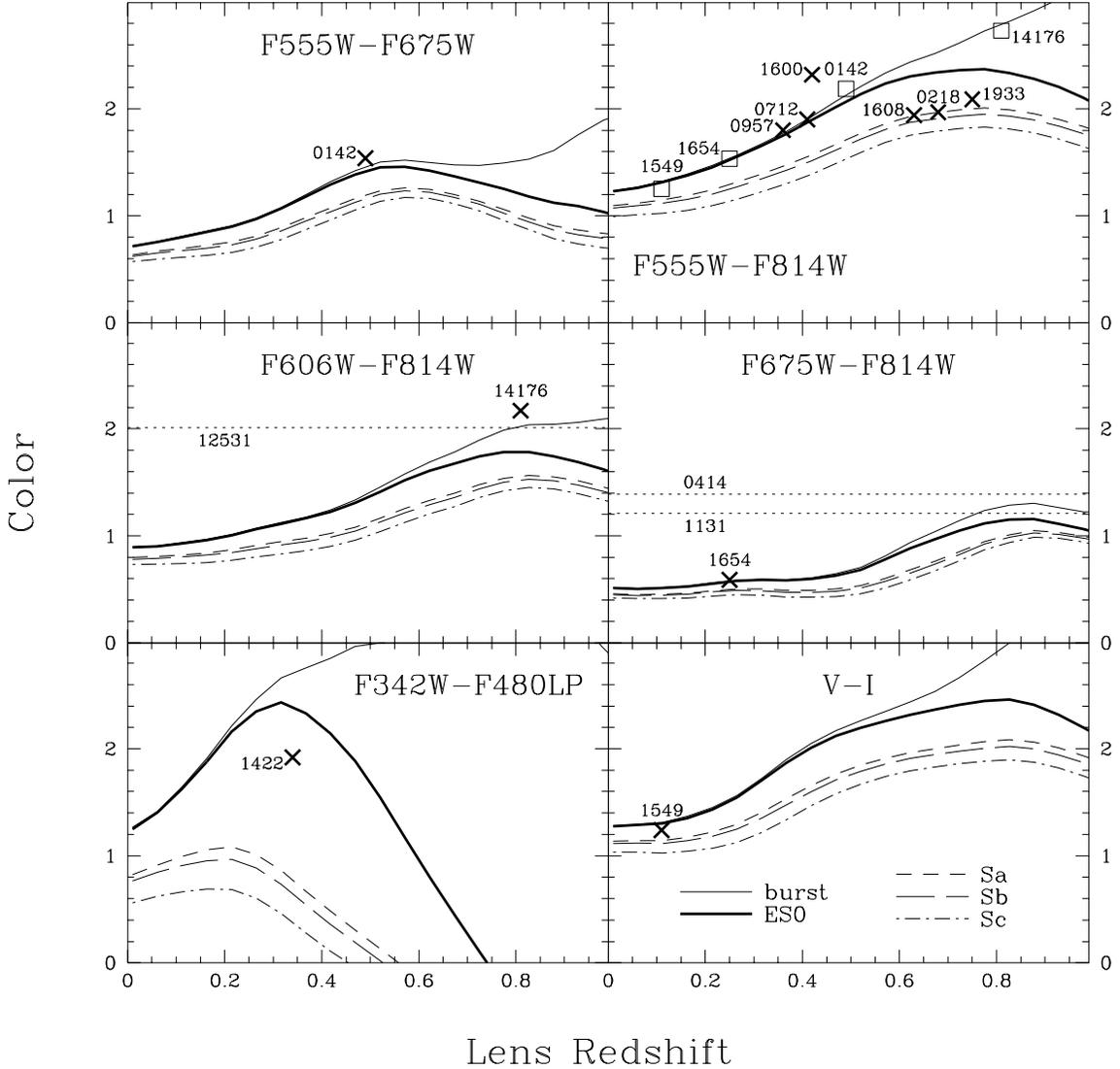}
	\caption{
Observed extinction-corrected colors and theoretical color evolution
curves.  Lens galaxies with a known redshift are shown as crosses,
while those with no known redshift are shown as horizontal dotted
lines.  For 0142, 14176, 1549, and 1654, we estimated F555W$-$F814W
colors by applying color transformations to the observed colors;
these points are shown as boxes.  Theoretical color evolution
curves are shown for various galaxy types:  the ``burst'' model
consists of a 1 Gyr period of constant star formation followed
by passive evolution; the E/S0 model has an exponential star
formation rate with a time scale of 1 Gyr; and the spiral models
have a star formation rate proportional to the gas fraction,
where the proportionality constant decreases from Sa to Sb to
Sc (see Guiderdoni \& Rocca-Volmerange 1988).  The burst and
E/S0 models have a Salpeter (1955) IMF, while the spiral models
have a Scalo (1986) IMF.  The curves shown are for $\Omega_0=1$
and $H_0=50$ \Hunits, with a galaxy formation redshift $z_f=15$.
}\label{fig:colors}
\end{figure}

\begin{figure}
	\plotone{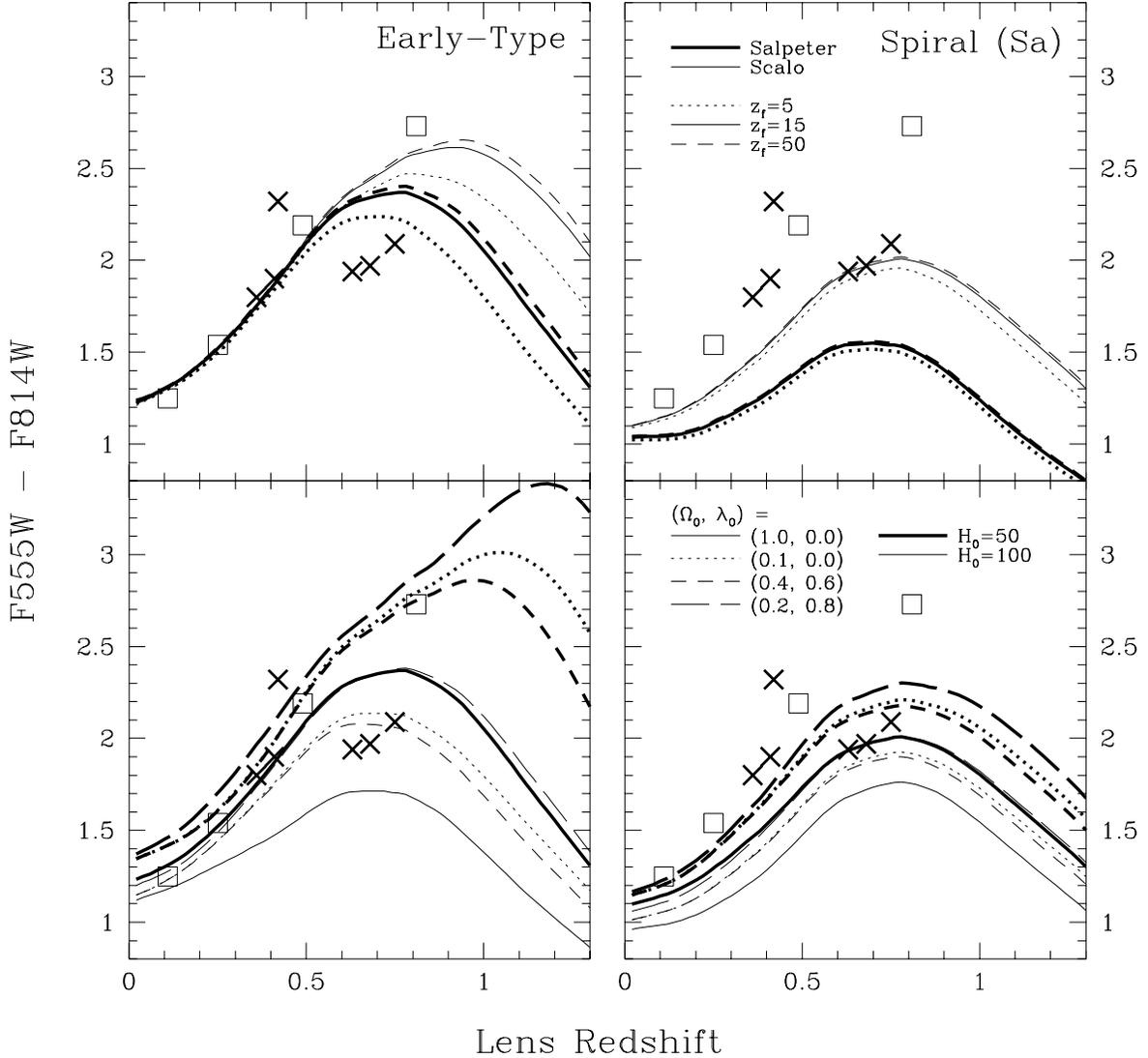}
	\caption{
Systematic effects in the theoretical color evolution curves
for early-type (left) and spiral (right) galaxy models.  In
all panels the crosses and boxes denote observed or estimated
extinction-corrected F555W$-$F814W colors taken from Figure~4.
{\it Top}:  The effects of changing the galaxy evolution model.
The heavy curves have a Salpeter (1955) IMF while the light
curves have a Scalo (1986) IMF.  The dotted, solid, and dashed
lines have a formation redshift $z_f=5$, $15$, and $50$,
respectively.  The cosmological model is $\Omega_0 = 1$ and
$H_0 = 50$ \Hunits.
{\it Bottom}:  The effects of changing the cosmological model.
The heavy (light) curves have $H_0 = 50$ (100) \Hunits.  The
solid, dotted, and dashed curves have different values for
$\Omega_0$ and $\lambda_0$.  All curves have $z_f = 15$, so the
present galaxy ages are $12.8\, h_{50}^{-1}$ Gyr ($\Omega_0=1$),
$17.0\, h_{50}^{-1}$ Gyr ($\Omega_0=0.1$ or $\lambda_0=0.6$), and
$20.6\, h_{50}^{-1}$ Gyr ($\lambda_0=0.8$).
}\label{fig:magtest}
\end{figure}

The distribution of lens galaxy colors provides a way to type the
galaxies, to probe galaxy evolution, and to make photometric redshift
estimates.  Fourteen lens galaxies have measured colors (see Table~2);
\reffig{colors} compares them to theoretical color evolution curves
for various galaxy types.  To help compare the colors we have taken
the lenses with known lens redshifts and used the color corrections
to transform the observed colors to F555W$-$F814W (roughly $V-I$),
as shown in \reffig{colors} (top right) and \reffig{magtest}.  In
\reffig{magtest} we also indicate the systematic effects in the
theoretical color evolution curves related to the IMF, formation
redshift, star formation rate, and cosmological model (see \S2).

Of the 14 measured colors, 11 are either consistent with or redder
than early-type galaxy models.  For B~1600$+$434 the red color
is not reliable because it is based on low signal-to-noise data;
Jaunsen \& Hjorth (1997) suggested from earlier ground-based
observations that the color and morphology of the lens galaxy in
B~1600$+$434 are actually more consistent with a spiral galaxy.
For other cases (MG~0414$+$0534, MG~1131$+$0456, and HST~12531$-$2914),
the significance of the colors depends heavily on the unknown lens
redshifts.  If they have $z_l \gtrsim 0.8$ their colors are only
modestly redder than those of passively evolving early-types.  By
contrast, if they are at $z_l \lesssim 0.8$ their colors suggest
that a sizable fraction of early-type galaxies must contain enough
dust to significantly affect their colors; Lawrence et al.\ (1995)
have suggested that this may be the case at least for MG~0414$+$0534.
The possibility of dust in early-type galaxies would have dramatic
consequences for galaxy evolution models, which usually assume that
early-types have little dust.  It would also affect cosmological
constraints derived from optically-selected lens statistics (see
Kochanek 1996a; Falco et al.\ 1997a; Malhotra et al.\ 1997).

The three lenses with colors bluer than those of early-type models
are B~0218$+$357, B~1608$+$656, and B~1933$+$503.  For B~1608$+$656
the color is hard to interpret due to strong contamination from the
lensed images (especially in $I$) and to the complicated structure
of the galaxy; the model fits give a redder color than the aperture
photometry, and neither is very reliable at present.  For B~1933$+$503
the color is based on very low signal-to-noise data and thus is not
very reliable.  By contrast, for B~0218$+$357 the blue color seems
reliable and suggestive of an Sa galaxy, which is consistent with
other observations.  B~0218$+$357 is the smallest-separation lens
known, with an image separation $0\parcs34$ that is typical of
expectations for lensing by spirals ($\langle\Delta\theta\rangle
\sim 0\parcs6$, see \S3.3), and the lens contains HI and molecular
gas (Carilli, Rupen \& Yanny 1993; Wiklind \& Combes 1995; Combes
\& Wiklind 1997), all of which provides strong evidence that the
lens galaxy is a spiral.

\reffig{magtest} illustrates the effects of varying the stellar
population and evolution models in interpreting the colors.  Most
lens galaxy colors are roughly consistent with passively evolving
early-type galaxies, and not with spiral galaxies, regardless of
the parameter choices.  The anomalous colors do not weaken this
conclusion, because varying the parameters in the plausible
direction (decreasing the formation redshift from $z_f=15$, or
increasing the Hubble constant from $H_0=50$ \Hunits) tends to
make the theoretical colors bluer rather than redder, so a color
that is difficult to explain as an early-type galaxy is even more
difficult to explain as a spiral galaxy.

\subsection{Image separations and lens luminosities}

Theoretical models predict correlations between lens luminosities
and image separations, which are used in statistical studies to
estimate the cosmological model (e.g.\ Turner et al.\ 1984;
Fukugita \& Turner 1991; Kochanek 1993a, 1996a; Maoz \& Rix 1993;
Falco et al.\ 1997a; Im, Griffiths \& Ratnatunga 1997).  Singular
isothermal sphere lens models relate the image separation
$\Delta\theta$ to the dark matter velocity dispersion $\sigma$ of
the lens galaxy by
$\Delta\theta/\Delta\theta_* = (\sigma/\sigma_*)^2 (D_{ls}/D_{s})$,
where $\Delta\theta_* = 8\pi(\sigma_*/c)^2$ is the image separation
produced by an $L_*$ galaxy for a source at infinity.  The average
observed separation is $\langle\Delta\theta\rangle \simeq
\Delta\theta_*/2$ (see Kochanek 1993b for general relations).
The Faber-Jackson (1976) and Tully-Fisher (1977) relations then
relate the lens galaxy's velocity dispersion to its luminosity
via $L/L_* = (\sigma/\sigma_*)^\gamma$, so we expect an image
separation/lens luminosity correlation of the form
\begin{eqnarray}
	{L \over L_*} &=& \left(
		{\Delta\theta \over \Delta\theta_*}\, {D_s \over D_{ls}}
	\right)^{\gamma/2} , \quad\mbox{or} \\
	M_B &=& M_{B*} - 1.25\,\gamma\,\log\left(
		{\Delta\theta \over \Delta\theta_*}\, {D_s \over D_{ls}}
	\right) .
	\label{eq:dthmag}
\end{eqnarray}
For early-type galaxies, Kochanek's (1993a, 1996a) estimate of the
Faber-Jackson relation based on gravitational lens statistics is
consistent with $\gamma=4$ and yields $\sigma_* = 220\pm20$ \kms,
which also agrees with dark matter models for the stellar dynamics
of ellipticals (Kochanek 1994).  For spiral galaxies, Fukugita \&
Turner's (1991) estimate of the Tully-Fisher relation is $\gamma
\approx 2.6$ and $\sigma_* = 144_{-13}^{+8}$ \kms.  These velocity
dispersions yield characteristic image splittings of
$\Delta\theta_*(\mbox{E/S0}) = 2\parcs79$ and
$\Delta\theta_*(\mbox{spiral}) = 1\parcs19$.
In other words, for the same luminosity a spiral produces a smaller
image separation than an early-type; conversely, for a fixed image
separation a spiral must be $\sim$2 mag brighter than an early-type.
Finally, from galaxy number counts and dynamical models, an $L_*$
galaxy has an absolute $B$ magnitude of $M_{B*} = (-19.7\pm0.1)+
5\log h$ (Efstathiou, Ellis \& Peterson 1988).

\begin{figure}
	\plotone{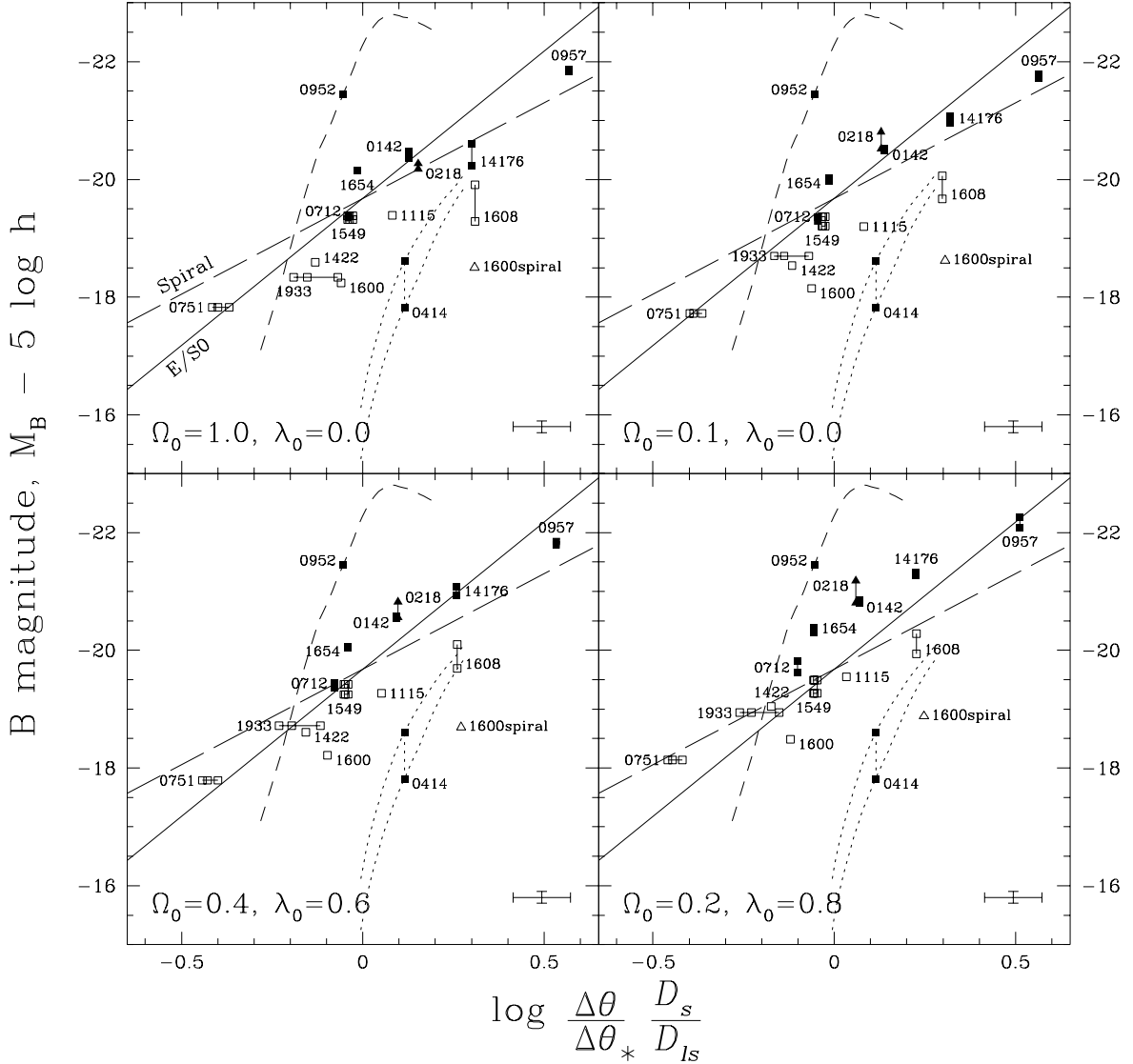}
	\caption{
Image separation/lens luminosity correlations.  The horizontal axis
is the normalized separation using $\Delta\theta = 2b_{SIS}$ with
$b_{SIS}$ from Table~5.  The vertical axis is the rest-frame
absolute $B$ magnitude using total magnitudes from Table~3 with
color, $K$, evolutionary, and Galactic extinction corrections.
Filled points denote plausible total magnitudes and open points
denote aperture magnitudes.  Triangles denote spirals and squares
denote early-types.  Where the lens redshift is unknown, the
points indicate the median expected redshift and the lines indicate
the $90\%$ confidence interval (see Table~1).  Where the source
redshift is unknown, points along the horizontal line denote
$z_s=2$, $3$, and $4$ (from right to left).  Lenses with
magnitudes in multiple bands are represented by points connected
vertically.  The expected relations for E/S0 and spiral galaxies
are shown by the solid and dashed lines.  The error bars indicate
the uncertainties in $\Delta\theta_*$ and $M_{B*}$.
}\label{fig:dthmag}
\end{figure}

\reffig{dthmag} shows the predicted relations for E/S0 and
spiral galaxies, together with the empirical results for lenses
with at least one known redshift.  The image separations were
taken to be $\Delta\theta = 2b_{SIS}$ with $b_{SIS}$ from Table~5.
The rest-frame $B$ luminosities were estimated by using the total
magnitudes in Table~3 with the color, $K$, evolutionary, and
Galactic extinction corrections described in \S2.  Some of the
magnitudes in Table~3 are estimates of the total integrated
magnitude, and we indicate these with filled points in
\reffig{dthmag}.  Other magnitudes are only aperture magnitudes,
which we indicate with open points.  Note that we have included
B~1600$+$434 both as an early-type and a spiral galaxy; the main
difference is the value of $\Delta\theta_*$ used in the
normalization of the image separation.  In general, the data show
the expected correlation, although there are a few lenses that
need special discussion:
\begin{itemize}

\item MG~0414$+$0534:  As with conclusions about the lens galaxy's
color (\S3.2), conclusions about its luminosity/image separation
relation depend strongly on the lens redshift.  If the lens has
$z_l \gtrsim 0.8$ (the upper end of the dotted curves in
\reffig{dthmag}), its color, luminosity, and image splitting are
all roughly consistent with passively evolving early-type models.
Note that without the correction for Galactic extinction, the
lens galaxy would appear anomalously red and faint even at
$z_l \gtrsim 0.8$.

\item B~1600$+$434:  Jaunsen \& Hjorth (1997) proposed from
ground-based images that the lens galaxy is a spiral, and the
HST images reveal a structure that is difficult to interpret but
is not inconsistent with a spiral (Jackson et al.\  1997).
However, \reffig{dthmag} shows that when treated as a spiral the
lens is underluminous by $\sim2$ mag, while when treated as an
early-type it sits significantly closer to the trend.  Its color
(see \S3.2) and low luminosity seem anomalous for a spiral, but
they are hard to interpret given the existing low signal-to-noise
data.

\item B~1608$+$656 and B~1933$+$503:  The colors, although very
uncertain, seemed to be more consistent with spiral galaxy models
than with early-type models (\S3.2).  The luminosities and image
separations, however, are consistent with the early-type
correlation and not with the spiral correlation.  Better optical
observations (including infrared observations) and lens models
are needed to properly interpret these systems.

\item MG~0751$+$2716, PG~1115$+$080, and B~1422$+$231:  Each
galaxy is part of a small group that contributes $\sim10\%$ to
the image separation (Hogg \& Blandford 1994; Keeton \& Kochanek
1997a; Kundi\'c et al.\ 1997ab; Leh\'ar et al.\ 1997; Schechter
et al.\ 1997; Tonry 1997), and thus is expected to appear somewhat
underluminous.  With total magnitude estimates and a quantitative
estimate of the correlation, it would be possible to see whether
this is true.

\item Q~0957$+$561:  The galaxy is part of a cluster that
contributes significantly to the large image separation (Young
et al.\ 1980), so its position below the trend is expected.
With a good quantitative estimate of the image separation/lens
luminosity correlation it would be possible to estimate the
intrinsic image splitting of the galaxy, and thus to break the
cluster degeneracy in lens models and estimates of the Hubble
constant (see Grogin \& Narayan 1996; Falco et al.\ 1997c;
Fischer et al.\ 1997).
\end{itemize}

Table~6 gives the parameters $M_{B*}$ and $\gamma$ derived from
the correlations by fitting a subsample of lens galaxies, namely
early-type galaxies with a known lens redshift (see the caption
of Table~6 for a list).  For each fit, a lens with both source
and lens redshifts and a magnitude in only one passband was
represented by a single data point.  A lens with no source
redshift or with magnitudes in multiple passbands was represented
by two data points that cover the range of results from different
passbands and/or different source redshifts ($2 \le z_s \le 4$).
Rather than trying to estimate the uncertainties, we assumed
uniform uncertainties scaled so that $\chi^2 = N_{dof}$ at the
minimum.  With this technique, the error bars give a rough estimate
of the uncertainties from the observed scatter.  The empirical
correlations are at least broadly consistent with a Faber-Jackson
coefficient $\gamma \approx 4$, although with the current data
the values are slightly lower ($\gamma \simeq 2.7\pm0.5$ for
$\Omega_0=1$).  The values for $M_{B*}$ are also lower than the
canonical value, but only by a few tenths of a magnitude.

Im et al.\ (1997) recently used the theoretical relation (\refeq{dthmag})
with a sample of seven lens galaxies to constrain the cosmological
model.  They found that lens galaxy luminosities are significantly
lower than expected unless $\lambda_0$ is large, which contradicts
results from lens statistics (Maoz \& Rix 1993; Kochanek 1996a; Falco
et al.\ 1997a) that rule out a large $\lambda_0$.  We believe that
the results of Im et al.\ (1997) are biased in two ways.  First, Im
et al.\ (1997) underestimated some of the lens galaxy magnitudes.
For lenses with only aperture magnitudes, Im et al.\ (1997)
subtracted $0.3$ mag to estimate total magnitudes; we believe this
to be a significant underestimate of the aperture corrections (compare
Tables~3 and 4).  In addition, Im et al.\ (1997) did not account for
Galactic extinction, which is significant for several lenses (see
Table~1).  By using underestimated magnitudes, Im et al.\ (1997)
forced the cosmological model to compensate by raising $\lambda_0$
to increase the luminosity inferred from a given apparent magnitude.
Second, Im et al.\ (1997) did not properly treat
the uncertainties in global parameters such as $M_{B*}$.  For example,
for a canonical value of $M_{B*}$ with uncertainty $\sigma_M$, and a
fitted value that differed by $\Delta M$, Im et al.\ (1997) would
have assigned $\Delta\chi^2 \simeq N_{lens} (\Delta M/\sigma_M)^2$
whereas a proper treatment of the covariance matrix would give
$\Delta\chi^2 \simeq (\Delta M/\sigma_M)^2$.  As a result, Im et
al.'s (1997) cosmological uncertainties were underestimated by a
factor of $2$--$3$, independent of the systematic biases from
underestimating the lens luminosities.  Taking into account the
underestimated luminosities and treating the uncertainties correctly
overwhelms any cosmological conclusions.  The image separation/lens
luminosity correlation certainly promises to give an interesting
cosmological constraint, but only after the quality of the data has
improved.

\subsection{Mass-to-light ratios and galaxy evolution}

Finally, we combined the masses from Table~5 with the aperture
magnitudes from Table~4 to compute mass-to-light ratios.  $M/L$
estimates from lensing have a unique advantage over estimates from
dynamical methods: whereas the dynamical methods are plagued by the
ambiguities of using stellar dynamical models to estimate the mass,
the lensing method has a robust mass estimate and thus is limited
almost entirely by the quality of the photometry.  In addition,
because mass-to-light ratios depend only on aperture magnitudes and
not on accurate profile fits and extrapolations, they yield physical
results that are more reliable than the total luminosities studied
in \S3.3.  As a result, lensing mass-to-light ratios offer an
excellent probe of galaxy evolution and structure.  Lensing also
makes it possible to compute a reliable $M/L$ estimate for individual
galaxies, as opposed to $M/L$ estimates from the fundamental plane
that are usually interpreted only in a statistical sense.

\begin{figure}
	\plotone{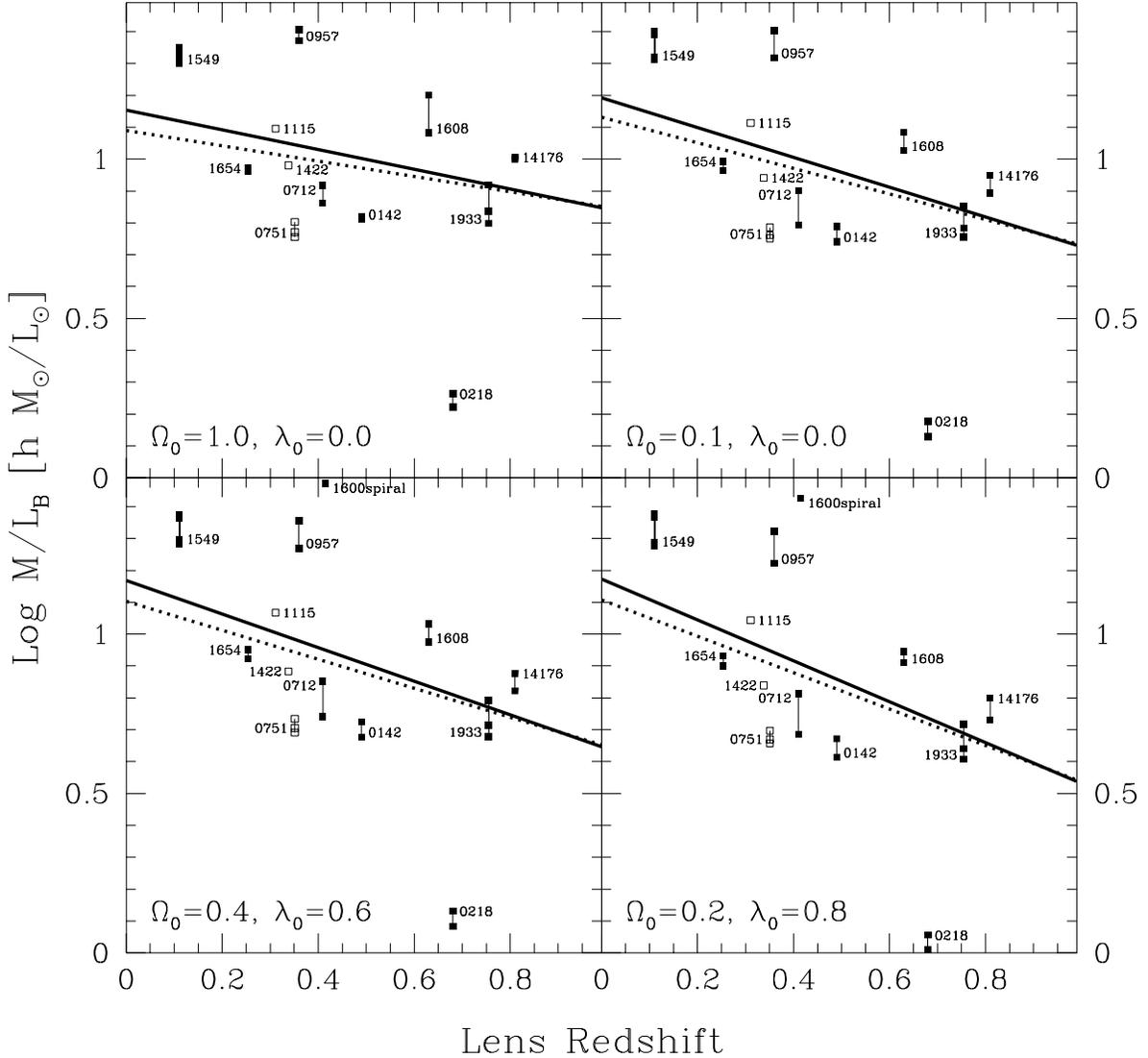}
	\caption{
Rest-frame $B$-band mass-to-light ratios inside the critical radius.
The luminosities were computed from the aperture magnitudes in
Table~4 with Galactic extinction, color, and $K$ corrections -- {\it
but no evolutionary corrections}.  Filled points indicate that the
aperture magnitudes used an aperture radius equal to the critical
radius, while open points indicate an aperture radius not equal to
the critical radius.  Vertical lines connect values of $M/L$
computed from different photometric bands, and (for 0751, 1549, and
1933) with different source redshifts ($z_s=2$, 3, and 4); thus the
spread in the points gives some idea of the systematic uncertainties.
0218 stands out because it is a spiral galaxy, and 0957 because the
mass from the cluster increases the $M/L$; 1600 is off the scale
in the two upper panels.  The clear trend of $M/L$ with redshift is
consistent with passive evolution.  The heavy lines show the best
fits to this trend; the solid lines exclude 0751, 1115, and 1422
from the fits, while the dotted lines include them (see the text
and Table~6).
}\label{fig:MLevol}
\end{figure}

\reffig{MLevol} shows the rest-frame $B$-band mass-to-light ratio
inside the critical radius for the lens galaxies with a known lens
redshift.  These $M/L_B$ use luminosities that were converted to
rest-frame $B$ by using the spectral evolution model to compute
color and $K$ corrections; we did {\it not\/} apply the evolutionary
correction because we wanted to look for evidence of evolution.
Note that although we are using $V$, $R$, and $I$ magnitudes to
compute the rest-frame $B$ luminosity, we are not actually
extrapolating the magnitudes; since rest-frame $B$ is roughly
equivalent to $V$ at $z \simeq 0.2$ and $I$ at $z \simeq 1$,
we are in fact interpolating between magnitudes.

Although there is significant scatter in \reffig{MLevol}, there
is a clear decrease in $M/L_B$ with redshift due to galaxy evolution.
To lowest order, $\log M/L_B(z)$ is expected to be linear in $z$,
\begin{equation}
	\log M/L_B(z) = \log M/L_B(0) + z [d(\log M/L_B)/dz] + \ldots
\end{equation}
(e.g.\ van Dokkum \& Franx 1996), so we can fit a line to the
data.  For the fit we excluded B~0218$+$357 because it is a spiral
galaxy, B~1600$+$434 because it is poorly understood, and Q~0957$+$561
because the cluster mass makes the $M/L$ something other than a true
galaxy $M/L$.  We used the same fit technique as in \S3.3, with
at most two data points per lens to represent the range of results,
and uniform errors scaled to give $\chi^2 = N_{dof}$ at the minimum.
Again the quoted error bars give a rough estimate of the uncertainties.

Table~6 summarizes the fits for the different cosmological models,
and \reffig{MLevol} shows the best fits superposed on the data.  We
note that of the lens galaxies used in the fit, only MG~0751$+$2716,
PG~1115$+$080, and B~1422$+$231 lack estimates of the magnitude
within the critical radius; we consider the fits both with and
without these galaxies.  Without these three galaxies, we find
$d(\log M/L_B)/dz = -0.3 \pm 0.1$ for $\Omega_0=1$ and $-0.5 \pm 0.1$
for $\Omega_0=0.1$ (with $\lambda_0=0$); including these galaxies
gives evolution rates that are $1\sigma$ smaller.  There is an
additional systematic uncertainty because in this preliminary
study we neglected the fact that $M/L$ has a small dependence on
luminosity (e.g.\ van der Marel 1991; J{\o}rgensen, Franx \&
Kj{\ae}rgaard 1996), and that with a dark matter lens model the
$M/L$ depends on the impact parameter.  Our results are comparable
to recent results from studies of intermediate-redshift clusters.
Kelson et al.\ (1997) found $d(\log M/L_V)/dz \sim -0.3$
($\Omega_0=0.1$) based on fundamental-plane observations of
clusters at redshifts of 0.33, 0.39, and 0.58; and Schade et al.\
(1997) found $d(\log L)/dz \sim 0.3$ ($\Omega_0=1$) based on a
projection of the fundamental plane in clusters at redshifts
between 0.2 and 1.  The fundamental plane studies are confined
to galaxies in high-density clusters, but Stanford et al.\ (1997)
suggest that evolution rates are not very sensitive to cluster
properties such as optical richness or X-ray luminosity.  The
fact that we find similar evolution rates for lens galaxies in
low-density environments provides further empirical evidence
that there is no strong environmental dependence in the stellar
populations of early-type galaxies.

\begin{figure}
	\plotone{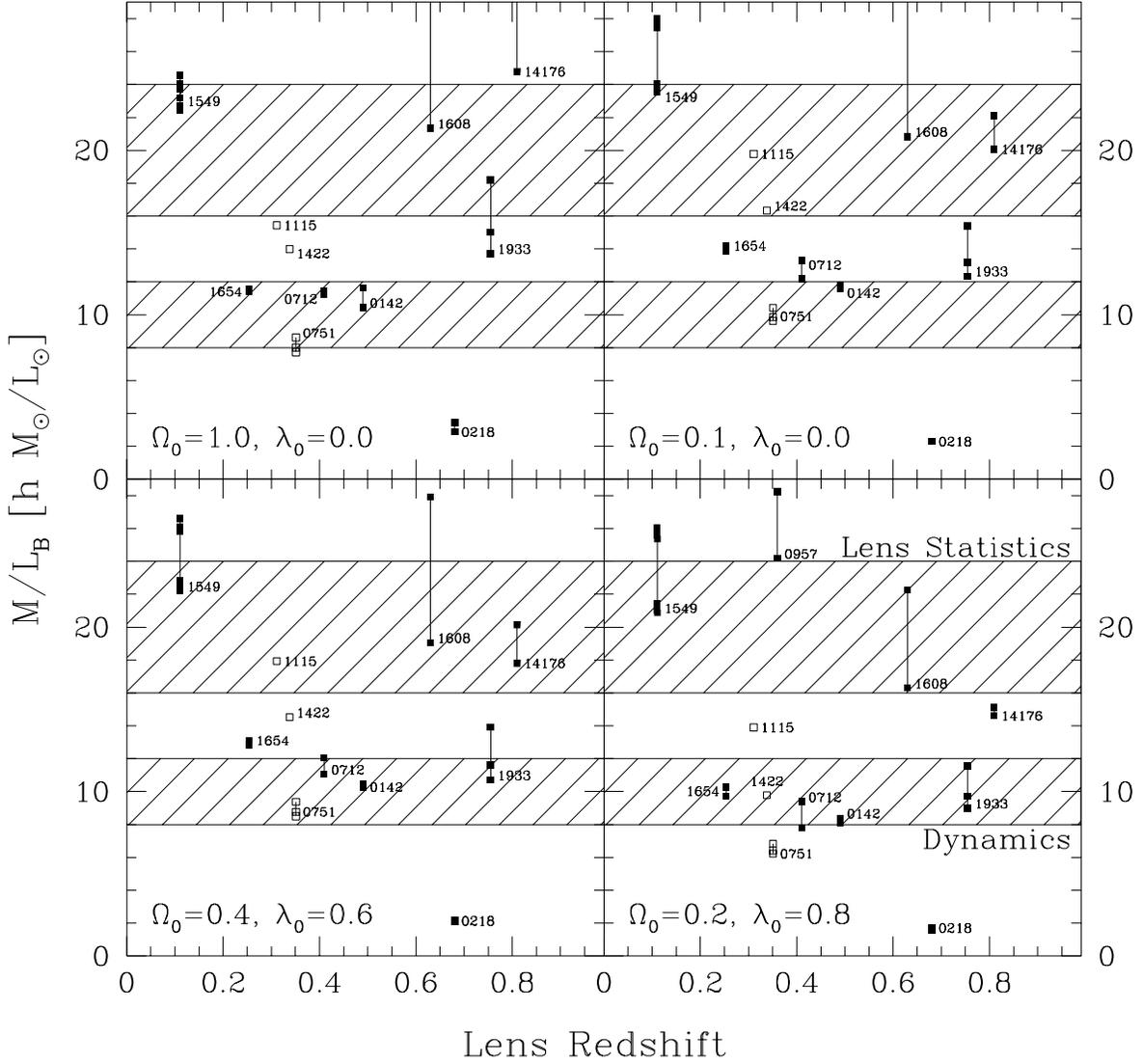}
	\caption{
Evolution-corrected $M/L$, i.e.\ the same $M/L$ as in Figure 7
except that the evolutionary correction has been added.  The
notation is the same as in Figure 7.  Again 0218 stands out
because it is a spiral, and 0957 (which is off the scale in all
but the $\lambda_0=0.8$ panel) because the mass from the cluster
increases the $M/L$.  The shaded regions indicate the results
from lens statistics (e.g.\ Maoz \& Rix 1993; Kochanek 1996a)
and from constant $M/L$ stellar dynamical models (e.g.\ van der
Marel 1991).
}\label{fig:ML}
\end{figure}

Mass-to-light ratios are also important in setting the mass scale
for both stellar dynamics and gravitational lensing.  Previous
studies have shown that consistency with constant $M/L$ stellar
dynamical models requires an average $M/L_B \sim (10\pm2)h$ (e.g.\
van der Marel 1991), while consistency with lens statistics requires
$M/L_{B*} = (20\pm4)h$ (e.g.\ Maoz \& Rix 1993; Kochanek 1996a),
and estimates from photometry and lens models of a few lenses have
found $M/L_B \sim 20h$ (e.g.\ Kochanek 1995; Burke et al.\ 1992).
\reffig{ML} shows the evolution-corrected $M/L_B$ (i.e.\ where the
rest-frame $B$ luminosity has been estimated by applying the
evolutionary correction in addition to the color, $K$, and Galactic
extinction corrections) for our larger sample.  Two lenses that
have notably low values are B~0218$+$357, which is a spiral and
has $M/L_B \sim 4h$ characteristic of spirals (e.g.\ Kent 1987;
Broeils \& Courteau 1997), and MG~0751$+$2716, which has only an
isophotal magnitude that overestimates the luminosity within the
critical radius and hence underestimates $M/L_B$.  Other than
these two lenses, the mass-to-light ratios are generally
higher than expected from stellar dynamics, although they are
somewhat lower than expected from lens statistics.  The large
scatter may be related to the dependence of $M/L$ on luminosity
and on impact parameter.  With better photometric data and a
larger lens sample it would be possible to correct for these
effects to reduce the scatter, and thus to see whether they
can explain why $M/L$ from lens models might be smaller than
those from lens statistics.  Nevertheless, even at present the
lens galaxy $M/L$ provide weak evidence for dark matter in the
inner regions of galaxies.

\section{Conclusions}

The sample of gravitational lens galaxies is now large enough
to warrant a systematic study of their properties.  By
combining the available optical data, including new surface
photometry of HST images, with constraints from lens models,
we have surveyed the physical properties of 17 lens galaxies at
redshifts between 0.1 and 0.8.  Most of the galaxies appear to
be passively evolving early-type galaxies, with the exception
of one clear spiral (B~0218$+$357).  Several lens galaxies
(B~1600$+$434, B~1608$+$656, and B~1933$+$503) have poor-quality
images or contamination from the lensed images and thus require
further study to understand their colors and morphologies;
nevertheless, at least for B~1608$+$656 and B~1933$+$503 the
image separations, lens luminosities, and mass-to-light ratios
are more consistent with early-type galaxies than with spirals.
Several other lens galaxies (MG~0414$+$0534, MG~1131$+$0456, and
HST~12531$-$2914) may or may not be anomalously red, depending
on their redshifts.  Our results do not support the suggestion
by Malhotra et al.\ (1997) that massive galaxies at intermediate
redshifts are very dusty, or by Jackson et al.\ (1997) that most
of the lenses are spirals.  Note that accepting either of these
hypotheses would force a radical revision of almost all galaxy
evolution models.

The shape distribution of lens galaxies is similar to that of
the general population of early-type galaxies, although there
is weak evidence for a deficit of apparently round lens galaxies;
such a deficit may be created by the inclination dependence of
lensing cross sections that bias flattened lens galaxies toward
being viewed edge-on (Keeton \& Kochanek 1997b).  There is no
obvious correlation between the optical and model ellipticities
of lens galaxies.  However, there is a strong correlation between
the optical and model position angles, suggesting that the
projected mass and projected light are generally aligned to
$\lesssim$10\deg, except in the presence of strong external
tidal perturbations.  This conclusion rules out dark halos that
are far out of equilibrium and have intrinsic axes misaligned
with respect to the luminous baryons.  It also suggests that
halos cannot be formed solely by dissipationless collapse,
because such halos tend to be very flat and nearly prolate
(e.g.\ Dubinski \& Carlberg 1991; Warren et al.\ 1992) so that,
when they are combined with a typical modestly triaxial or oblate
luminous galaxy (e.g.\ Franx et al.\ 1991), there is a large
misalignment between the major axes of the projected mass and
light (Romanowsky \& Kochanek 1997).  The interaction of the
dark matter with the baryons must substantially alter the shape
of the dark matter halo so that the light and mass have similar
triaxialities and intrinsic axes, as is seen in the preliminary
simulations of Dubinski (1994).

Lens galaxies obey the correlations between image separation and
lens luminosity predicted by dark matter lens models combined with
the Faber-Jackson and Tully-Fisher relations.  For the early-type
lens galaxies in an $\Omega_0=1$ cosmology, the characteristic
magnitude is $M_{B*} = (-19.3\pm0.1) + 5\log h$ and the
``Faber-Jackson'' exponent is $\gamma = 2.7\pm0.5$, with an
additional systematic uncertainty due to uncertainties in total
magnitude estimates, and to the uncertainty in $\sigma_*$ from
lens statistics (e.g.\ Fukugita \& Turner 1991; Kochanek 1993a,
1996a).  Im et al.\ (1997) attempted to use the image separation/lens
luminosity correlation to constrain the cosmological model, but
by underestimating aperture corrections, neglecting Galactic
extinction, and improperly treating uncertainties they obtained
misleading results favoring a high-$\lambda_0$ cosmology.  A
robust calculation of the image separation/lens luminosity
correlation would offer an interesting cosmological constraint,
but at present the data lack the necessary precision.  The
correlation would also improve the constraints on $H_0$ by
allowing a calculation of the intrinsic image splitting of lens
galaxies to break the cluster degeneracy in Q~0957$+$561 (see
Grogin \& Narayan 1996; Falco et al.\ 1997c; Fischer et al.\
1997) and the group degeneracy in PG~1115$+$080 (see Schechter
et al.\ 1997; Keeton \& Kochanek 1997a; Courbin et al.\ 1997b),
but again the required precision is not yet available.  It
might also be possible to search for the analog of the
fundamental plane as an explanation of the scatter in the
correlation, but the current sample has too few robust
estimates of effective radii.

The most robust physical properties of the lens galaxies that we
can calculate are mass-to-light ratios.  Mass-to-light ratios
require only aperture magnitudes, so they do not depend on accurate
profile fits and extrapolations.  In addition, lensing measures a
mass that has an internal uncertainty of only a few percent (e.g.\
Kochanek 1991a; Wambsganss \& Paczy\'nski 1994) and systematic
uncertainties of $\lesssim$20\% (due to external tidal perturbations,
potential fluctuations associated with large-scale structure, or
the cosmological model; e.g.\ Bar-Kana 1996; Wambsganss et al.\
1997; Keeton et al.\ 1997).  Thus lens galaxy mass-to-light ratios
are limited primarily by the quality of the photometry, making them
an outstanding probe of galaxy evolution.  We measure an evolution
rate of $d(\log M/L_B)/dz = -0.3\pm0.1$ ($-0.5\pm0.1$) for
$\Omega_0=1$ ($0.1$), although there is an additional systematic
uncertainty $M/L$ depends weakly on luminosity and more strongly
on impact parameter in dark matter models, and in this preliminary
study we did not include corrections.  Our results for lens
galaxies in low-density environments are comparable to results
from measurements of the fundamental plane in intermediate-redshift
clusters (e.g.\ Kelson et al.\ 1997; Schade et al.\ 1996, 1997),
suggesting that there are no strong environmental effects in the
evolution of early-type galaxies (see also Stanford et al.\ 1997).
The evolution-corrected mass-to-light ratios help distinguish
between early-type lens galaxies (which have $M/L_B \sim 10$--$20h$,
e.g.\ van der Marel 1991; Maoz \& Rix 1993; Kochanek 1996a) and
spiral lens galaxies (which have $M/L_B \sim 4h$, e.g.\ Kent 1987;
Broeils \& Courteau 1997); at present, B~0218$+$357 is the only
lens galaxy with a robust $M/L_B$ estimate that is consistent with
a spiral galaxy.  For the early-type lens galaxies, the mass-to-light
ratios are generally larger than expected from constant $M/L$ stellar
dynamical models, although the scatter is large.  Most of the
scatter is due to uncertainties and systematic effects in the
photometric data; in particular, it may be related to the dependence
of $M/L$ on luminosity and impact parameter.  With improved
photometric data the uncertainties would be significantly reduced,
and lens galaxy mass-to-light ratios could provide strong evidence
for dark matter in the inner parts of galaxies.

Our analysis is limited primarily by the quality of the optical data
and by the absence of redshift measurements for some of the lens
systems.  Given a homogeneous data set with well-determined photometry,
most of the observational uncertainties will be eliminated.  Then by
using its ability to probe mass distributions and measure masses --
and thus to avoid difficult spectroscopy and dynamical analysis of
distant galaxies -- we can use gravitational lensing as a powerful
probe of high-redshift galaxies and their evolution.  In addition,
with well-calibrated correlations it should be possible to use lens
galaxy colors, luminosities, scale lengths, and image separations
as an accurate method for estimating lens redshifts.

\acknowledgements

Acknowledgements:  
We thank J.~Leh\'ar, B.~McLeod, E.~Woods, and J.~Huchra for useful
discussions; in particular, J.~Leh\'ar and B.~McLeod assisted with
the analysis of several lens systems.
Support for this work was provided by the Smithsonian Institution,
and by NASA through grant number 05505 from the Space Telescope
Science Institute, which is operated by the Association of
Universities for Research in Astronomy, Inc., under NASA contract
NAS5-26555.
CRK is supported by ONR-NDSEG grant N00014-93-I-0774.
CSK is supported by NSF grant AST-9401722 and NASA ATP grant NAG5-4062.



\clearpage

\begin{deluxetable}{rcccll}
\tablewidth{0pt}
\tablecaption{Gravitational Lens Geometries, Galactic Extinctions, and Redshifts}
\tablehead{
 \colhead{Name} &\colhead{Geometry\tm{\S}} &\colhead{$E(B-V)$\tm{\#}} &\colhead{$z_s$} &\colhead{$z_l$} &\colhead{Reference} 
}
\startdata
   Q~0142$-$100\0 & Double (O)         &$ 0.05 $&$ 2.72      $&$ 0.49$                  & Su87              \\
   B~0218$+$357\0 & Double+Ring (R, O) &$ 0.10 $&$ 0.96      $&$ 0.68$                  & Pa93, Br93, Lw96  \\
  MG~0414$+$0534  & Quad+Arc (R, O)    &$ 0.18 $&$ 2.64      $&$ 0.44_{0.15}^{0.85}\tdd$& He92, Lw95        \\
   B~0712$+$472\0 & Quad (R, O)        &$ 0.15 $&$ 1.33      $&$ 0.41$                  & Br97              \\
  MG~0751$+$2716  & Ring (R)           &$ 0.07 $&             &$ 0.35$                  & Le97              \\
 BRI~0952$-$0115  & Double (O)         &$ 0.07 $&$ 4.5\0     $&$ 0.99_{0.31}^{1.96}\tdd$& Mm92              \\
   Q~0957$+$561\0 & Double (R, O)      &$ 0.01 $&$ 1.41      $&$ 0.36$                  & Wa79, Yo80        \\
  PG~1115$+$080\0 & Quad (O)           &$ 0.06 $&$ 1.72      $&$ 0.31$                  & We80, Ku97a, To97 \\
  MG~1131$+$0456  & Ring+Arc (R, O)    &$ 0.06 $&             &                         & He88              \\
HST~12531$-$2914  & Quad (O)           &$ 0.10 $&             &                         & Ra95              \\
HST~14176$+$5226  & Quad (O)           &$ 0.02 $&$(3.4)\td\0 $&$ 0.81$                  & Ra95, Cr96        \\
   B~1422$+$231\0 & Quad (R, O)        &$ 0.04 $&$ 3.62      $&$ 0.34$                  & Pa92, Ku97b, To97 \\
  MG~1549$+$3047  & Ring (R)           &$ 0.04 $&             &$ 0.11$                  & Le93              \\
   B~1600$+$434\0 & Double (R, O)      &$ 0.02 $&$ 1.57      $&$ 0.41$                  & Jc95, Br97        \\
   B~1608$+$656\0 & Quad+Arc (R, O)    &$ 0.05 $&$ 1.39      $&$ 0.63$                  & My95, Fs96        \\
  MG~1654$+$1346  & Ring (R)           &$ 0.09 $&$ 1.74      $&$ 0.25$                  & Ln89              \\
   B~1933$+$503\0 & Quad\tm{**}\ \ (R) &$ 0.16 $&             &$ 0.76$                  & Br97              \\
\tableline
\enddata
\tablenotetext{\S}{Ring indicates a ring of lensed extended radio emission,
and arc indicates lensed extended optical emission.  Double or quad indicates
two or four images.  R and O indicate whether the lensed images have been
detected at radio and optical wavelengths.}
\tablenotetext{**}{B~1933$+$503 has a complicated geometry with as many as
10 images.  It appears to consist of three sources, two of which are quadruply
imaged and one of which is doubly imaged (Browne et al.\ 1997).}
\tablenotetext{\#}{Galactic extinction, in magnitudes, computed by
estimating the HI column density $N_H$ from Stark et al.\ (1992) and then
converting to $E(B-V)$ using $N_H/E(B-V)=5.9\times10^{21}\mbox{ mag}^{-1}
\mbox{ cm}^{-2}$ from Spitzer (1978).}
\tablenotetext{\dag}{$(\cdots)$ denotes a tentative measurement of the
source redshift.}
\tablenotetext{\ddag}{Estimated median lens redshift and $90\%$ confidence
interval (for an $\Omega_0=1$ cosmology), computed from the probability of
producing the observed image separation (see Kochanek 1992).}
\end{deluxetable}

\def\pha{\phantom{0\pm0.00}}

\begin{deluxetable}{rccllcl}
\tablewidth{0pt}
\tablecaption{Lens Galaxy Structural Parameters and Colors}
\small
\tablehead{
\colhead{Name} &\colhead{$R_e$ (arcsec)} &\colhead{$e$} &\colhead{PA ($^\circ$)} &\colhead{Filters} &\colhead{Color} &\colhead{Method, Ref}
}
\startdata
   Q~0142$-$100\0 & $(0.50 \pm 0.03) $ &$ 0.31 \pm0.03  $&\phs$ 63\pm4$ & F555W, F675W  &$ 1.57  $& B,   new         \\
   B~0218$+$357\0 &                    &                 &              & F555W, F814W  &$ 2.10  $& A,   Jc97, Ml97  \\
  MG~0414$+$0534  & $ 1.28 \pm 0.09  $ &$ 0.29 \pm0.04  $&\phs$ 81\pm4$ & F675W, F814W  &$ 1.51  $& B,   Fl97        \\
   B~0712$+$472\0 & $(0.42 \pm 0.04) $ &$ 0.59 \pm0.03  $&\phs$ 60\pm2$ & F555W, F814W  &$ 2.09  $& B,   Jc97        \\
  MG~0751$+$2716  & $(0.27 \pm 0.03) $ &$ 0.49 \pm0.03  $&\phs$ 17\pm2$ & R             &         & A,   Le97        \\
 BRI~0952$-$0115  & $(0.14 \pm 0.03) $ &$ 0.43 \pm0.08  $&\phs$ 61\pm7$ & F675W         &         & B,   new         \\
   Q~0957$+$561\0 & $ 4.63 \pm 0.07  $ &$ 0.21 \pm0.01  $&\phs$ 49\pm2$ & F555W, F814W  &$ 1.82  $& B,   Be97        \\
  PG~1115$+$080\0 &                    &                 &              & F785LP        &         &      Kr93        \\
  MG~1131$+$0456  & $ 0.90 \pm 0.08  $ &$ 0.12 \pm0.04  $&\phs$ 36\pm10$& F675W, F814W  &$ 1.25  $& B,   new         \\
HST~12531$-$2914  & $ 0.19 \pm 0.02  $ &$ 0.17 \pm0.05  $&\phs$ 20\pm8$ & F606W, F814W  &$ 2.11  $& B,   Ra95        \\
HST~14176$+$5226  & $ 1.13 \pm 0.05  $ &$ 0.30 \pm0.02  $&\phs$ 37\pm2$ & F606W, F814W  &$ 2.19  $& B,   Ra95        \\
   B~1422$+$231\0 & $(0.8\0\pm 0.2\0)$ &$ 0.27 \pm0.13  $&    $-59\pm15$& F342W, F480LP &$(2.0)\0$& A,   Im96        \\
  MG~1549$+$3047  & $(3.5)\pha       $ &$ 0.35 \pm0.05  $&    $-40\pm5$ & V, I          &$(1.3)\0$& A,   Le93, Le96  \\
   B~1600$+$434\0 &                    &$(0.4\0\pm0.1\0)$&\phs$ 45\pm5$ & F555W, F814W  &$(2.35) $& B/C, Jc97        \\
   B~1608$+$656\0 & $(0.39 \pm 0.04) $ &$ 0.60 \pm0.03  $&\phs$ 81\pm2$ & F555W, F814W  &$(2.00) $& B/C, Jc97        \\
  MG~1654$+$1346  & $ 1.80 \pm 0.02  $ &$ 0.40 \pm0.01  $&    $-83\pm1$ & F675W, F814W  &$ 0.65  $& B,   new         \\
   B~1933$+$503\0 &                    &$(0.57 \pm0.03) $&    $-41\pm2$ & F555W, F814W  &$(2.30) $& B/C, Jc97        \\
\tableline
\enddata
\normalsize
\tablecomments{
Each lens galaxy is described by its effective radius $R_e$, ellipticity
$e=1-b/a$, major axis position angle, and color.  The colors have not
been corrected for Galactic extinction.  The methods are as follows:
(A) Results taken from the cited literature.
(B) Results determined from an elliptical de Vaucouleurs profile fit
to HST images; results based on archival images cite the first known
publication of the observations, and results based on our new
observations are listed as ``new.''
(C) Color determined inside an $0\parcs3$ radius aperture centered on
the galaxy.
Error bars are standard errors using a $\chi^2$ rescaled to be unity per
degree of freedom at the minimum.  Formal uncertainties for the colors from
profile fits are negligible, but systematic uncertainties are probably
$\sim$0.2 mag.  $(\cdots)$ denotes a value that is systematically uncertain
because the galaxy is too faint or its light is significantly affected by
the lensed images (see text).  Structural parameters left blank were
unavailable in the literature and/or impossible to determine reliably from
HST images; colors left blank indicate that observations were available in
only one passband.
}
\end{deluxetable}

\begin{deluxetable}{rcccrcrcc}
\tablewidth{0pt}
\tablecaption{Lens Galaxy Total Magnitudes}
\small
\tablehead{
 \colhead{Name} & \colhead{F814W} &\colhead{F675W} &\colhead{F555W} & \multicolumn{4}{c}{Other} &\colhead{Method}
}
\startdata

    Q~0142$-$100\0 &          &$ 19.26 $&$ 20.83  $&          &          &         &        & B \\
    B~0218$+$357\0 &$(20.0)\0$&         &$(22.0)\0$&          &          &         &        & A \\
   MG~0414$+$0534  &$ 20.50  $&$ 22.01 $&          &          &          &         &        & B \\
    B~0712$+$472\0 &$ 19.48  $&         &$ 21.57  $&          &          &         &        & B \\
   MG~0751$+$2716  &          &         &          &      (R) &$[21.3]\0$&         &        & A \\
  BRI~0952$-$0115  &          &$ 21.89 $&          &          &          &         &        & B \\
    Q~0957$+$561\0 &$ 16.43  $&         &$ 18.24  $&          &          &         &        & B \\
   PG~1115$+$080\0 &          &         &          & (F785LP) &$[18.4]\0$&         &        & A \\
   MG~1131$+$0456  &$ 20.76  $&$ 22.01 $&          &          &          &         &        & B \\
 HST~12531$-$2914  &$ 21.42  $&         &          &  (F606W) &$ 23.54  $&         &        & B \\
 HST~14176$+$5226  &$ 19.53  $&         &          &  (F606W) &$ 21.72  $&         &        & B \\
    B~1422$+$231\0 &          &         &          & (F480LP) &$[21.6]\0$& (F342W) &$[23.6]$& A \\
   MG~1549$+$3047  &          &         &          &      (I) &$[16.3]\0$&     (V) &$[17.6]$& A \\
    B~1600$+$434\0 &$[20.4]\0$&         &          &          &          &         &        & B \\
    B~1608$+$656\0 &$[19.7]\0$&         &$[22.7]\0$&          &          &         &        & B \\
   MG~1654$+$1346  &$ 17.46  $&$ 18.10 $&          &          &          &         &        & B \\
    B~1933$+$503\0 &$[21.9]\0$&         &          &          &          &         &        & B \\
\tableline
\enddata
\normalsize
\tablecomments{
The magnitudes have not been corrected for Galactic extinction.
Method A: magnitudes taken from the literature.
Method B: total magnitudes determined from elliptical de Vaucouleurs profile
fits.
Formal uncertainties for the magnitudes from profile fits are negligible, but
systematic uncertainties are probably $\sim$0.2 mag.  $(\cdots)$ denotes a
magnitude that is uncertain by an estimated $0.5$ mag, while $[\cdots]$ denotes
a magnitude that has no aperture correction and thus underestimates the total
magnitude.  The references are the same as in Table~2.
}
\end{deluxetable}

\def\pha{\phantom{[}}
\def\phb{\phantom{]}}

\begin{deluxetable}{rcccrcrcc}
\tablewidth{0pt}
\tablecaption{Lens Galaxy Aperture Magnitudes}
\small
\tablehead{
 \colhead{Name} & \colhead{F814W} &\colhead{F675W} &\colhead{F555W} & \multicolumn{4}{c}{Other} &\colhead{Method}
}
\startdata
    Q~0142$-$100\0 &           &$ 19.54 $&$ 21.11  $&          &          &         &         & B \\
    B~0218$+$357\0 & $(21.4)\0$&         &$(23.7)\0$&          &          &         &         & A \\
   MG~0414$+$0534  & $ 21.19  $&$ 22.70 $&          &          &          &         &         & B \\
    B~0712$+$472\0 & $ 19.82  $&         &$ 21.90  $&          &          &         &         & B \\
   MG~0751$+$2716  &           &         &          &      (R) &$[21.3]\0$&         &         & A \\
  BRI~0952$-$0115  &           &$ 22.03 $&          &          &          &         &         & B \\
    Q~0957$+$561\0 & $ 17.35  $&         &$ 19.17  $&          &          &         &         & B \\
   PG~1115$+$080\0 &           &         &          & (F785LP) &$[18.4]\0$&         &         & A \\
   MG~1131$+$0456  & $ 21.45  $&$ 22.70 $&          &          &          &         &         & B \\
 HST~12531$-$2914  & $ 21.67  $&         &          &  (F606W) &$ 23.77  $&         &         & B \\
 HST~14176$+$5226  & $ 20.06  $&         &          &  (F606W) &$ 22.24  $&         &         & B \\
    B~1422$+$231\0 &           &         &          & (F480LP) &$[21.6]\0$& (F342W) &$[23.6]$ & A \\
   MG~1549$+$3047  &           &         &          &      (I) &$(17.8)\0$&     (V) &$(19.1)$ & A \\
    B~1600$+$434\0 & $(21.4)\0$&         &$(23.9)\0$&          &          &         &         & B \\
    B~1608$+$656\0 & $(19.7)\0$&         &$(22.7)\0$&          &          &         &         & B \\
   MG~1654$+$1346  & $ 18.44  $&$ 19.09 $&          &          &          &         &         & B \\
    B~1933$+$503\0 & $(21.9)\0$&         &          &          &          &         &         & B \\
\tableline
\enddata
\normalsize
\tablecomments{
Magnitudes measured inside a circular aperture with radius equal to the critical radius
($b_{SIS}$) of the lens models in Table 5.  They have not been corrected for Galactic
extinction.
Method A: magnitudes taken from the literature.
Method B: magnitudes determined from elliptical de Vaucouleurs profile fits.
Formal uncertainties for the magnitudes from profile fits are negligible, but
systematic uncertainties are probably $\sim$0.2 mag.  $(\cdots)$ denotes a
magnitude that is uncertain by an estimated $0.5$ mag, while $[\cdots]$ denotes
an aperture magnitude whose aperture radius is not equal to the critical radius.
The references are the same as in Table~2.
}
\end{deluxetable}

\def\ta{\tablenotemark{a}}
\def\tb{\tablenotemark{b}}
\def\tc{\tablenotemark{c}}
\def\td{\tablenotemark{d}}

\begin{deluxetable}{rlllllrl}
\tablecaption{Singular Isothermal Lens Models}
\tiny
\tablehead{
\colhead{Name} &\colhead{$b_{SIS}$ ($''$)} &\colhead{$\gamma$} &\colhead{$b_{SIE}$ ($''$)} &\colhead{$e$} &\colhead{PA ($^\circ$)\tm{\dag}} &\colhead{$\chi^2/N_{dof}$\tm{\ddag}} &\colhead{Ref}
}
\startdata
    Q~0142$-$100\0  &$1.17\pm0.05$ &$0.08\pm0.02$ & $1.14\pm0.02$ &$0.21\pm0.05$ &\phs$ 76_{-18}^{+9} $ &$(0)\0\0  0/0$ & HST\ta     \\
    B~0218$+$357\0  &$0.17       $ &$           $ & $           $ &$           $ &    $               $ &               & Pa95\td    \\
   MG~0414$+$0534   &$1.18\pm0.03$ &$0.10\pm0.03$ & $1.14\pm0.03$ &$0.38\pm0.09$ &\phs$ 79\pm1        $ &$(116)  111/6$ & Ka97       \\
    B~0712$+$472\0  &$0.69\pm0.02$ &$0.05\pm0.04$ & $0.69\pm0.01$ &$0.25\pm0.11$ &\phs$ 50\pm1        $ &$(28)\0  24/6$ & HST        \\
   MG~0751$+$2716   &$0.40       $ &$0.09       $ & $0.41       $ &$0.34       $ &\phs$ 64            $ &               & Le97a\tb   \\
  BRI~0952$-$0115   &$0.52\pm0.01$ &$0.07\pm0.01$ & $0.51\pm0.01$ &$0.19\pm0.03$ &\phs$ 65\pm5        $ &$(0)\0\0  0/0$ & HST        \\
    Q~0957$+$561\0  &$           $ &$           $ & $3.09       $ &$0.64       $ &\phs$ 69\pm1        $ &               & Le97b\tb   \\
   PG~1115$+$080\0  &$1.14\pm0.01$ &$0.12\pm0.02$ & $1.08\pm0.03$ &$0.48\pm0.07$ &\phs$ 67\pm1        $ &$(250)  483/6$ & Co97       \\
   MG~1131$+$0456   &$0.92\pm0.01$ &$0.11\pm0.01$ & $0.92       $ &$0.33       $ &    $-26            $ &               & Ch95\tb, Ch93 \\
 HST~12531$-$2914   &$0.55\pm0.03$ &$0.15\pm0.05$ & $0.54\pm0.04$ &$0.38\pm0.17$ &\phs$ 19\pm4        $ &$(20)\0  35/6$ & Ra95       \\
 HST~14176$+$5226   &$1.42\pm0.06$ &$0.15\pm0.04$ & $1.34\pm0.09$ &$0.52\pm0.11$ &\phs$ 49\pm3        $ &$(96)   111/6$ & Ra95       \\
    B~1422$+$231\0  &$0.77\pm0.01$ &$0.26\pm0.01$ & $0.65\pm0.02$ &$0.63\pm0.03$ &    $-53\pm1        $ &$(40)   124/6$ & Pa92, Im96 \\
   MG~1549$+$3047   &$           $ &$           $ & $1.15       $ &$0.07       $ &    $-48            $ &               & Le93\tb    \\
    B~1600$+$434\0  &$0.70       $ &$           $ & $           $ &$           $ &    $               $ &               & Jc95\td    \\
    B~1608$+$656\0  &$1.10\pm0.03$ &$0.06\pm0.04$ & $1.07\pm0.04$ &$0.35\pm0.16$ &\phs$ 69\pm2        $ &$(837)  790/6$ & HST        \\
   MG~1654$+$1346   &$0.98\pm0.01$ &$0.08\pm0.01$ & $0.98       $ &$0.27       $ &    $-81            $ &               & Ko95\tb, Ln90 \\
    B~1933$+$503\0  &$0.50\pm0.01$ &$0.15\pm0.02$ & $0.45\pm0.01$ &$0.60\pm0.03$ &    $-46\pm1        $ &$(10)\0\0 3/4$ & Ma97\tc    \\
\tableline
\enddata
\small
\tablenotetext{\dag}{Major-axis PA for the SIE lens models.  The PAs for
the SIS+shear models are the same to within the error bars.}
\tablenotetext{\ddag}{The $\chi^2$ for the SIS+shear and SIE models, and
the number of degrees of freedom, in the form:  ($\chi^2_{SIS}$) $\chi^2_{SIE}$
/ $N_{dof}$.}
\tablenotetext{a}{HST denotes a model based on data from our analysis of the
HST images.}
\tablenotetext{b}{Model results taken from the literature.}
\tablenotetext{c}{The $\chi^2$ for B~1933$+$503 is deceptively low.  The
image position error bars $\sigma_p$ were not given in Marlow et al.\
(1997), so actually $\chi^2_{SIE} = 3 (0\parcs01/\sigma_p)^2$ (and similar
for $\chi^2_{SIS}$).  In addition, the models used only the quadruply-imaged
flat spectrum source and neglected the other lensed images.}
\tablenotetext{d}{Critical radius estimated from a singular isothermal
sphere lens model to produce the observed image separation.}
\tablecomments{
Results from singular isothermal sphere plus external shear (SIS+shear;
$b_{SIS}$ and shear $\gamma$) and singular isothermal ellipsoid (SIE;
$b_{SIE}$ and ellipticity $e$) lens models, based on data from
the cited literature.  Point-image lenses were modeled by fitting the
quasar image positions and flux ratios as well as the galaxy position
if available (e.g.\ Keeton \& Kochanek 1997a).  Radio rings were modeled
using the LensClean program (e.g.\ Kochanek 1995; Chen et al.\ 1995).
The error bars are $1\sigma$ standard errors using a $\chi^2$ renormalized
to equal $N_{dof}$ at the minimum; because the models generally are
poor fits, these error bars overestimate the mass uncertainties.  The
SIS model gives a robust estimate for the mass within the critical
radius, $M=\pi b^2 \Sigma_{cr}$, which depends only weakly on the
lens model.  The ellipticity depends on the radial mass profile of
the lens, roughly as $(1-\kappa_r)$ where $\kappa_r$ is the surface
density of the model at the critical radius in units of the critical
surface density -- more centrally concentrated models require higher
ellipticities.  The position angle is essentially model-independent
in single-shear models.
}
\normalsize
\end{deluxetable}

\def\td{\tablenotemark{\dag}}
\def\tdd{\tablenotemark{\ddag}}

\begin{deluxetable}{ccccc}
\tablewidth{0pt}
\tablecaption{Empirical Results}
\tablehead{
 \colhead{$(\Omega_0,\lambda_0)$} &\colhead{$M_{B*} - 5 \log h$\td} &\colhead{$\gamma$\td} &\colhead{$\log M/L_B(0)$\tdd} &\colhead{$d(\log M/L_B)/dz$\tdd}
}
\startdata
$(1.0,0.0)$ & $-19.3\pm0.1$ & $2.7\pm0.5$ & $(1.09)$ $1.15\pm0.04$ & $(-0.24)$ $-0.31\pm0.08$ \\
$(0.1,0.0)$ & $-19.4\pm0.1$ & $3.3\pm0.4$ & $(1.13)$ $1.19\pm0.04$ & $(-0.40)$ $-0.47\pm0.08$ \\
$(0.4,0.6)$ & $-19.6\pm0.1$ & $3.2\pm0.4$ & $(1.10)$ $1.17\pm0.05$ & $(-0.45)$ $-0.53\pm0.08$ \\
$(0.2,0.8)$ & $-20.0\pm0.1$ & $3.7\pm0.5$ & $(1.11)$ $1.17\pm0.05$ & $(-0.57)$ $-0.64\pm0.08$ \\
\tableline
\enddata
\tablenotetext{\dag}{
Fits to the lens luminosity/image separation correlations in \S3.3, using
the early-type lens galaxies with a known lens redshift (0142, 0712, 0751,
1115, 14176, 1422, 1549, 1608, 1654, and 1933).  Including 0957 in the fits
gives results that are statistically consistent with those in the table.
Error bars were computed assuming uniform errors scaled so that $\chi^2 =
N_{dof}$ at the minimum.
}
\tablenotetext{\ddag}{
Fits to the mass-to-light ratio evolution in \S3.4.  The results in
parentheses use all of the early-type lens galaxies with a known lens
redshift (listed above).  The other results exclude 0751, 1115, and 1422
because these lens galaxies lack accurate estimates of the magnitude
within the critical radius.  Again error bars were computed assuming
uniform uncertainties scaled so that $\chi^2 = N_{dof}$ at the minimum.
}
\end{deluxetable}


\clearpage

\end{document}